\documentclass[aps,pra,twocolumn,superscriptaddress,floatfix,nofootinbib,longbibliography]{revtex4-2}

\usepackage{amsmath,braket,dsfont,verbatim,latexsym,amssymb,indentfirst,mathrsfs,mathtools,amsthm,bbm,bm,url,cancel,enumitem,setspace}
\usepackage[colorlinks=true,linkcolor=blue,bookmarks=false,citecolor=magenta,urlcolor=blue]{hyperref}
\usepackage[caption=false]{subfig}

\usepackage{verbatim,indentfirst}
\usepackage[title,titletoc]{appendix}
\usepackage{graphicx}
\usepackage{booktabs}
\usepackage[dvipsnames]{xcolor}
\usepackage[capitalise]{cleveref}

\newcommand{\A}{\texttt{Alice}}
\newcommand{\B}{\texttt{Bob}}
\newcommand{\E}{\texttt{Eve}}
\newcommand{\AB}{\texttt{Alice}\ and \texttt{Bob}}
\newcommand{\Y}{\texttt{Yves}}

\begin{document}

\title{Proof-of-Principle Implementation of Quantum Key Distribution in an Indefinite Causal Order}

\author{Yann Valibouse}
\email{yann.valibouse@univie.ac.at}
\affiliation{University of Vienna, Faculty of Physics, Vienna Center for Quantum Science and Technology (VCQ), Boltzmanngasse 5, 1090 Vienna, Austria}
\affiliation{University of Vienna, Vienna Doctoral School in Physics, Boltzmanngasse 5, 1090 Vienna, Austria}

\author{Martí Cladera-Rosselló}
\affiliation{University of Vienna, Faculty of Physics, Vienna Center for Quantum Science and Technology (VCQ), Boltzmanngasse 5, 1090 Vienna, Austria}
\affiliation{University of Vienna, Vienna Doctoral School in Physics, Boltzmanngasse 5, 1090 Vienna, Austria}

\author{Michael Antesberger}
\affiliation{University of Vienna, Faculty of Physics, Vienna Center for Quantum Science and Technology (VCQ), Boltzmanngasse 5, 1090 Vienna, Austria}
\affiliation{University of Vienna, Vienna Doctoral School in Physics, Boltzmanngasse 5, 1090 Vienna, Austria}

\author{Hector Spencer-Wood}
\affiliation{School of Physics and Astronomy, University of Glasgow, G12 8QQ Glasgow, United Kingdom}

\author{Kyrylo Simonov}
\affiliation{Fakultät für Mathematik, Universität Wien, Oskar-Morgenstern-Platz 1, 1090 Vienna, Austria}

\author{Patrik Sund}
\affiliation{University of Vienna, Faculty of Physics, Vienna Center for Quantum Science and Technology (VCQ), Boltzmanngasse 5, 1090 Vienna, Austria}

\author{Mathieu Bozzio}
\affiliation{University of Vienna, Faculty of Physics, Vienna Center for Quantum Science and Technology (VCQ), Boltzmanngasse 5, 1090 Vienna, Austria}

\author{Philip Walther}
\affiliation{University of Vienna, Faculty of Physics, Vienna Center for Quantum Science and Technology (VCQ), Boltzmanngasse 5, 1090 Vienna, Austria}
\affiliation{Christian Doppler Laboratory for Photonic Quantum Computer, Boltzmanngasse 5, 1090 Vienna, Austria}
\affiliation{Institute for Quantum Optics and Quantum Information (IQOQI) Vienna, Austrian Academy of Sciences, Boltzmanngasse 3, 1090 Vienna, Austria}

\author{Lee A. Rozema}
\affiliation{University of Vienna, Faculty of Physics, Vienna Center for Quantum Science and Technology (VCQ), Boltzmanngasse 5, 1090 Vienna, Austria}

\date{\today}

\begin{abstract}
In quantum physics, the order in which different operations occur can be placed in superposition. The resulting processes have an indefinite causal order, which is of fundamental interest and can be viewed as a novel quantum resource enabling new protocols. Here, we report an experimental proof-of-principle  implementation of one such protocol, where we perform BB84-like quantum cryptography by placing \AB's measurement-and-preparation operations in a photonic quantum SWITCH. By embedding \AB\ within the quantum SWITCH, the protocol achieves an average eavesdropper detection probability of $0.15 \pm 0.02$ per shared qubit, with eavesdropper detection performed through measurements of the control qubit rather than comparing the key. Unlike standard BB84 and related schemes, which detect eavesdropping by publicly revealing and discarding a fraction of the raw key, our approach requires no disclosure of key material: every retained qubit can, in principle, be tested for eavesdropping while remaining available for key generation.  Although the present implementation does not yet constitute a secure quantum key distribution protocol, owing to post-selection required for measurements within the quantum SWITCH, it provides a proof of principle that indefinite causal order can be exploited to detect eavesdropping without sacrificing key bits.

\end{abstract}

\maketitle
\section{Introduction}

Causal order plays a central role in our description of physical processes, where operations are typically assumed to occur in a well-defined temporal sequence. However, quantum theory and attempts to reconcile it with the dynamical nature of spacetime have motivated the exploration of scenarios in which causal order itself becomes a quantum degree of freedom. In such situations, different orders of quantum operations can be coherently superposed, giving rise to the phenomenon of \emph{indefinite causal order}
 (ICO)~\cite{Hardy_2007, Hardy2009, Baumeler, Wechs, oreshkov}. One of the most prominent realizations of ICO is the quantum SWITCH, a higher-order quantum operation that coherently controls the order in which two or more quantum operations are applied via a control quantum system (qubit)~\cite{Chiribella}. Rather than implementing a fixed causal structure, the quantum SWITCH realizes a coherent superposition of alternative orders.

The quantum SWITCH has been shown to provide operational advantages over causally ordered strategies in a wide range of quantum-information tasks, including communication through noisy channels~\cite{Feix_2015, Guerin, ebler2018enhanced, Procopio2019, Loizeau2020, goswami2020increasing, Caleffi2020, Mukhopadhyay2020, Chiribella2021Comm, Caleffi2023, wang2025exp} and metrology~\cite{zhao2020quantum, Bavaresco2021, Chapeau2021, Chapeau2022, metro, Liu2023, Chapeau2023, Goldberg2023}, along with applications in quantum computing~\cite{Araujo2014comp, Renner2022, EscandonMonardes2023, yin2024magic, simonov2026comp} and thermodynamics~\cite{felce2020quantum, Guha2020, simonov2022work, Goldberg2023thermo, Zhu2023, simonov2025activation, li2025enhancing, xue2025heat, lisboa2026, lai2026charging}. These results suggest that ICO could be a valuable resource for quantum technologies beyond its foundational significance.

The potential role of ICO in quantum cryptography has begun to attract attention only recently. In Ref.~\cite{spencer2025indefinite}, a quantum key distribution (QKD) protocol based on the quantum SWITCH was proposed, in which \AB's measurements are performed in a coherent superposition of causal orders. The central idea is that an eavesdropper acting inside the SWITCH disturbs the interference between the alternative orders, leaving a detectable signature in the control qubit. Consequently, the presence of an eavesdropper can be monitored through measurements of the control system, rather than through the public comparison of a subset of the raw key. More recently, ICO has also been shown to enhance communication-security-related figures of merit. In particular, Ref.~\cite{wu2025crypto} demonstrated that the quantum SWITCH can increase the private capacity of a specific QKD channel and, consequently, improve the maximum secret-key rate achievable under certain noise conditions. Furthermore, Ref.~\cite{lesniak2026} introduced a QKD protocol in which the secret key is generated from correlations that are incompatible with any definite causal order between \AB's operations. Together, these works highlight the emerging potential of ICO as a resource for secure quantum communication.

In this work, we present an experimental proof-of-principle realization of the protocol proposed in Ref.~\cite{spencer2025indefinite}. In the original QKD protocol, named BB84~\cite{bennett1984quantum}, as well as in representative prepare-and-measure~\cite{Scarani2009, Lo2005, Laing2010, Sidhu2025, LM05} and entanglement-based variants~\cite{E91, bbm92, PingPong}, the presence of an eavesdropper is estimated by publicly comparing a randomly selected subset of the raw key. While this procedure provides a reliable estimate of the quantum bit error rate, every disclosed bit must subsequently be discarded, creating an intrinsic trade-off between parameter estimation and key generation. This trade-off can be particularly stringent in scenarios where finite-size effects dominate~\cite{Tomamichel2012, Diamanti2016}. In contrast, the protocol considered here shifts eavesdropper detection from the key itself to the quantum degree of freedom that encodes the causal order of \AB. For the intercept-resend attacks studied in this work, an eavesdropper acting within the quantum SWITCH disrupts the interference between the alternative causal orders, leaving a measurable signature in the control system. As a result, the disturbance introduced by an eavesdropper can be inferred from measurements of the control qubit, without publicly revealing any portion of the raw key.

In the proposal of Ref.~\cite{spencer2025indefinite}, the quantum SWITCH is seen as a process modifying the quantum channel between \AB\ from the traditional BB84 protocol to a coherent superposition of causally ordered channels. In that sense, one photon entering the quantum SWITCH uses indefinite causal order as a resource to obtain more information per channel use (a key bit and a detection bit) than in BB84. The dimensionality of the protocol is therefore larger than that of a standard, polarization-encoded BB84 scheme. However, practically speaking, this extra dimensionality is not the crucial figure of merit limiting secure key rates. Our experiment shows that creating a larger system involving a target and a control qubit does not slow down the protocol's repetition rate. What matters is the number of repetition cycles of the protocol required to achieve a secure key of a given length. In this respect, our protocol is advantageous, as it allows for a reduction of up to a factor of $2$ in the number of repetition cycles needed to achieve a given key length, compared to a standard BB84 scheme.

A central experimental challenge in implementing the protocol of Ref.~\cite{spencer2025indefinite} is that both \AB\ must be able to access the shared key bit while preserving the coherence between the alternative causal orders. In conventional photonic implementations of the quantum SWITCH, measurement outcomes can generally only be accessed after the photon exits the quantum SWITCH. This prevents \AB\ from obtaining their measurement results locally within the quantum SWITCH, as required by the protocol considered here.

To overcome this limitation, we employ a recently developed measurement scheme that enables local polarization readout inside the quantum SWITCH through the use of a time-delocalized ancilla system~\cite{valibouse2026time}. The notion of time-delocalized systems and operations provides the appropriate framework for describing parties acting in an ICO~\cite{oreshkov2019time}. Operationally, an ancilla photon is prepared to interact with the system photon via entangling gates placed in each causal-order branch. These interactions would, in principle, encode which-order information in the ancilla and thereby destroy the coherence between the alternative causal orders. By erasing this which-order information through coherent recombination of the ancilla paths, the coherence of the quantum SWITCH is preserved, and the system photon's polarization, which encodes the key, can be read out locally via the ancilla polarization. The implementation relies on post-selected linear-optical interactions, which introduce practical security considerations for a fully deployable cryptographic system. Nevertheless, it provides a viable platform for experimentally investigating how ICO can be harnessed in quantum cryptographic protocols.

The remainder of the paper is organized as follows. Section~\ref{sec:theoretical_framework} introduces the theoretical framework underlying the quantum SWITCH and the QKD protocol based on it. Section~\ref{sec:setup} describes the photonic implementation of the quantum SWITCH and its adaptation to the proposed cryptographic scheme as a proof-of-principle. The experimental results, both in the absence and in the presence of an eavesdropper, are presented in Section~\ref{sec:results}. In Section~\ref{sec:conclusions}, we draw our conclusions.

\section{Theoretical Framework}
\label{sec:theoretical_framework}

\subsection{Quantum SWITCH}

There have been numerous experimental investigations of ICO over the past decade. The most widely studied realization is the quantum SWITCH, a higher-order quantum operation that coherently controls the order in which two or more quantum operations are applied to a target system $s$. The order of the operations is determined by an external quantum degree of freedom, known as the control system $c$, which can itself be prepared in a superposition state. In this way, the quantum SWITCH enables the target system to evolve in a coherent superposition of alternative causal orders. To date, the vast majority of experimental demonstrations have been performed on photonic platforms, where different degrees of freedom of a single photon encode the control and target systems~\cite{Procopio, Rubino, Wei, Guo, Rubino_comm, Rubino2022experimental, Cao, Cao_SDI, Zhu2023, MinAn, process, stromberg, rozema_review, richter}, although alternative implementations and proposals based on nuclear magnetic resonance and cavity quantum electrodynamics have also been discussed~\cite{nie2022experimental, procopio2025cavity}.

In the simplest scenario, the operations placed in a superposition of orders are two unitary gates $A$ and $B$ acting on $s$, while the causal order is controlled by a qubit $c$. The quantum SWITCH then implements the {operation}
\begin{align*}
    S(A,B)
    &= \ket{0}_c\!\bra{0} \otimes AB
     + \ket{1}_c\!\bra{1} \otimes BA \\
    &= \tfrac{1}{2}\,\mathbb{I}_c \otimes \{A,B\}
     + \tfrac{1}{2}\,\sigma_z^{(c)} \otimes [A,B],
\end{align*}
acting jointly on the control and target systems. Here, $\mathbb{I}_c$ and $\sigma_z^{(c)}$ denote the identity and Pauli-$Z$ operators acting on the control qubit, while $\{A,B\}=AB+BA$ and $[A,B]=AB-BA$ are the anticommutator and commutator of the two operations, respectively. The state of the control qubit determines the order in which the gates are applied: if the control is in the state $\ket{0}_c$, the target undergoes the sequence $AB$, whereas for $\ket{1}_c$ it undergoes the sequence $BA$. When the control is prepared in the superposition state $\omega_c=\ket{+}_c\!\bra{+}$, where $\ket{+}_c = \frac{1}{\sqrt{2}}(\ket{0}_c + \ket{1}_c)$, the target evolves in a coherent superposition of the two alternative orders $AB$ and $BA$.

More generally, let $\mathcal{A}$ and $\mathcal{B}$ be quantum channels described by the sets of Kraus operators $\{A_i\}_i$ and $\{B_j\}_j$, respectively. The quantum SWITCH transforms them into a new quantum channel acting on both $s$ and $c$ as
\begin{equation}
    \mathcal{S}(\mathcal{A}, \mathcal{B})[\omega_c \otimes \rho_s]
    = \sum_{i,j} S_{ij} \, (\omega_c \otimes \rho_s) \, S_{ij}^\dagger,
    \label{eq:S_channel}
\end{equation}
where
\begin{align*}
    S_{ij} &= \ket{0}_c\!\bra{0} \otimes A_i B_j
           + \ket{1}_c\!\bra{1} \otimes B_j A_i \\
           &= \tfrac{1}{2} \, \mathbb{I}_c \otimes \{A_i, B_j\}
            + \tfrac{1}{2} \, \sigma_z^{(c)} \otimes [A_i, B_j].
\end{align*}
\subsection{QKD Protocol and Its Correctness}

We now adapt the framework to a BB84-inspired quantum key distribution protocol. \AB\ independently choose a measurement basis labelled by $\mu\in\{0,1\}$, where $\mu = 0$ corresponds to the computational basis $\{\ket{0}_s, \ket{1}_s\}$, and $\mu = 1$ to the Hadamard basis $\{\ket{+}_s, \ket{-}_s\}$. For each basis, the possible measurement outcomes are labelled by $b \in \{0,1\}$. The corresponding rank-one projectors $P_b^{(\mu)}$ are given by
\begin{align} \label{eq:statesBB84}
    P_0^{(0)} &= \ket{0}_s\!\bra{0}, & P_1^{(0)} &= \ket{1}_s\!\bra{1}, \nonumber \\
    P_0^{(1)} &= \ket{+}_s\!\bra{+}, & P_1^{(1)} &= \ket{-}_s\!\bra{-}.
\end{align}
A measurement in basis $\mu$ evolves a state $\rho$ as
\begin{equation*}
    \mathcal M^{(\mu)}[\rho] = \sum_{b \in \{0,1\}} P_b^{(\mu)}\rho P_b^{(\mu)}.
\end{equation*}
Since the basis is chosen at random, the full local operation performed by each party before reconciliation is the random measurement channel
\begin{equation*}
    \mathcal M[\rho_s] = \sum_{\mu \in \{0,1\}} p_\mu \mathcal M^{(\mu)}[\rho_s] = \sum_{b, \mu \in \{0,1\}} M_b^{(\mu)}\rho_s M_b^{(\mu)\dagger},
\end{equation*}
where
\begin{equation*}
    M_b^{(\mu)} = \sqrt{p_\mu}\,P_b^{(\mu)}, \qquad p_0=p_1=\frac{1}{2}.
\end{equation*}
\AB\ perform the same random channel, hence, $\mathcal{A} = \mathcal{B} = \mathcal{M}$.

Following Equation~(\ref{eq:S_channel}), the quantum SWITCH of the two honest operations is therefore described by Kraus operators
\begin{align*}
    S_{b\mu,b'\mu'} &= \ket{0}_c\!\bra{0} \otimes M_b^{(\mu)}M_{b'}^{(\mu')} + \ket{1}_c\!\bra{1} \otimes M_{b'}^{(\mu')}M_b^{(\mu)} \\
    &= \tfrac{1}{2}\mathbb I_c\otimes \{M_b^{(\mu)},M_{b'}^{(\mu')}\} + \tfrac{1}{2}\sigma_z^{(c)}\otimes [M_b^{(\mu)},M_{b'}^{(\mu')}].
\end{align*}
After basis reconciliation, \AB\ retain only the rounds for which they selected the same basis, i.e., $\mu = \mu'$. On this reconciled subensemble,
\begin{equation*}
    M_b^{(\mu)}M_{b'}^{(\mu)} = p_\mu\,P_{b}^{(\mu)}P_{b'}^{(\mu)} = p_\mu\,\delta_{bb'}P_b^{(\mu)},
\end{equation*}
which implies
\begin{equation*}
    [M_b^{(\mu)},M_{b'}^{(\mu)}] = 0.
\end{equation*}
Therefore, for the retained rounds, the action of the quantum SWITCH is governed by Kraus operators
\begin{equation*}
    S_{b\mu,b'\mu} = p_\mu\,\delta_{bb'}\, \mathbb I_c\otimes P_b^{(\mu)}.
\end{equation*}
Hence, only events corresponding to identical outcomes ($b=b'$) contribute to the retained data, ensuring perfect correlations between \AB\ (see Figure~\ref{fig:protocol}).

Conditioning on a fixed reconciled basis $\mu$ and renormalizing the post-selected ensemble yields
\begin{equation*}
    \mathcal S_{\mu}(\mathcal M^{(\mu)},\mathcal M^{(\mu)}) [\omega_c\otimes\rho_s] = \omega_c\otimes \Bigl(\sum_{b \in \{0,1\}} P_b^{(\mu)}\rho_s P_b^{(\mu)}\Bigr).
\end{equation*}
Averaging uniformly over the two reconciled bases gives
\begin{equation}
    \mathcal S_{\rm rec}[\omega_c\otimes\rho_s] = \tfrac{1}{2} \omega_c\otimes \Bigl( \sum_{b,\mu \in \{0,1\}} P_b^{(\mu)} \rho_s P_b^{(\mu)} \Bigr).
\end{equation}
Thus, in the absence of an eavesdropper, \AB\ obtain perfectly correlated outcomes while the control qubit remains unchanged.

\subsection{Cheat Detection}

We now introduce an eavesdropper \E, described by a channel $\mathcal E$ with Kraus operators $\{E_k\}_k$ acting on $s$. For simplicity, we consider the case in which \E\ acts between \AB\ inside the quantum SWITCH. Restricting attention to the rounds that survive basis reconciliation, the relevant Kraus operators governing the action of the quantum SWITCH are
\begin{align*}
    S_{b\mu,b'\mu, k}^{(E)} &= p_\mu \Bigl(\ket{0}_c\!\bra{0} \otimes P_b^{(\mu)}E_kP_{b'}^{(\mu)} \\
    &\qquad + \ket{1}_c\!\bra{1} \otimes P_{b'}^{(\mu)}E_kP_b^{(\mu)}\Bigr).
\end{align*}
Introducing the generalized anticommutator and commutator
\begin{align*}
    \{P_b^{(\mu)},E_k,P_{b'}^{(\mu)}\} &:= P_b^{(\mu)}E_kP_{b'}^{(\mu)} + P_{b'}^{(\mu)}E_kP_b^{(\mu)}, \\
    [P_b^{(\mu)},E_k,P_{b'}^{(\mu)}] &:= P_b^{(\mu)}E_kP_{b'}^{(\mu)} - P_{b'}^{(\mu)}E_kP_b^{(\mu)},
\end{align*}
one obtains
\begin{align}\label{eq:EveKraus}
    S_{b\mu,b'\mu, k}^{(E)} &= \tfrac{1}{2}p_\mu \Bigl( \mathbb I_c \otimes \{P_b^{(\mu)},E_k,P_{b'}^{(\mu)}\} \nonumber \\
    & \qquad + \sigma_z^{(c)} \otimes [P_b^{(\mu)},E_k,P_{b'}^{(\mu)}] \Bigr).
\end{align}
Therefore, after conditioning and averaging over the retained bases, the resulting output state is
\begin{widetext}
\begin{align}
    \mathcal S_{\rm rec}^{(E)} [\omega_c\otimes\rho_s] &= \tfrac{1}{8} \sum_{b,b',\mu,k} \Big( \omega_c \otimes \{P_b^{(\mu)},E_k,P_{b'}^{(\mu)}\} \rho_s  \{P_b^{(\mu)},E_k,P_{b'}^{(\mu)}\}^{\dagger} \nonumber \\
    &\qquad + \sigma_z^{(c)} \omega_c \sigma_z^{(c)} \otimes [P_b^{(\mu)},E_k,P_{b'}^{(\mu)}] \rho_s [P_b^{(\mu)},E_k,P_{b'}^{(\mu)}]^{\dagger} \Big).
\end{align}
\end{widetext}
\noindent In particular, for an initial control state $\omega_c=\ket{+}_c\!\bra{+}$, the probability of obtaining the outcome $\ket{-}_c$ is
\begin{equation}
    p_- = \tfrac{1}{8} \sum_{b,b',\mu, k} \mathrm{Tr} \!\left[ [P_b^{(\mu)},E_k,P_{b'}^{(\mu)}] \rho_s [P_b^{(\mu)},E_k,P_{b'}^{(\mu)}]^{\dagger} \right].
\end{equation}
In the honest case, corresponding to $E_k\propto\mathbb I$, all generalized commutators vanish and therefore $p_- = 0$. In contrast, an eavesdropping operation can make
\begin{equation*}
    P_b^{(\mu)}E_kP_{b'}^{(\mu)} \neq P_{b'}^{(\mu)}E_kP_b^{(\mu)},
 \end{equation*}
resulting in a nonzero probability of obtaining the outcome $\ket{-}_c$.

In the honest protocol, the two branches interfere perfectly and the private control remains in the state $\ket{+}_c$. \E's intervention introduces distinguishability between the two orders, thereby reducing the interference visibility. This loss of coherence manifests itself as a nonzero population in the state $\ket{-}_c$, providing a direct signature of eavesdropping. Consequently, \A\ can detect \E\ without publicly revealing a subset of the raw key, unlike the usual BB84 parameter-estimation procedure. Once again, we emphasize that this is a convenient feature for lossy implementations that are sensitive to finite-size effects~\cite{Tomamichel2012,Diamanti2016}.

It should be noted that, unlike in the standard BB84 protocol, the present setting admits more than one possible location for an eavesdropper. In particular, Ref.~\cite{spencer2025indefinite} considered a more general attack with two eavesdroppers, \E\ and \Y, located at the two distinct positions available within the quantum SWITCH. \E\ acts between \AB's operations, while \Y\ acts at the second access point to the target system, outside the sequence of \AB's operations but still inside the quantum SWITCH. The two eavesdroppers can coordinate their actions through a correlated attack. In this case, the Kraus operators \eqref{eq:EveKraus} are replaced by
\begin{align*}
    S_{b\mu,b'\mu, k, l}^{(EY)} &= p_\mu \alpha_{kl} \Bigl(\ket{0}_c\!\bra{0} \otimes P_b^{(\mu)}E_kP_{b'}^{(\mu)}Y_l \\
    &\qquad + \ket{1}_c\!\bra{1} \otimes Y_l P_{b'}^{(\mu)}E_kP_b^{(\mu)}\Bigr),
\end{align*}
where  $\{ E_k \}_k$ and $\{Y_l\}_l$ are the Kraus operators of \E\ and \Y's operations, while $\alpha_{kl} \in \mathbb{C}$ are appropriately normalized coefficients. As shown in Ref.~\cite{spencer2025indefinite}, such attacks cannot extract information about the key without inducing a nonzero population in the control state $|-\rangle_c$. In what follows, however, we restrict ourselves to the simpler case of a single eavesdropper, \E.

\section{Experimental Setup}
\label{sec:setup}

We implement a proof-of-principle realization of the protocol using a photonic quantum SWITCH, in which the control and target qubits are encoded in different degrees of freedom of a single photon. The control qubit is encoded in the photon's path, while the target qubit is encoded in its polarization. The computational-basis states $\ket{0}$ and $\ket{1}$ are represented by horizontal ($\ket{H}$) and vertical ($\ket{V}$) polarization, respectively, whereas the Hadamard-basis states $\ket{+}$ and $\ket{-}$ correspond to diagonal ($\ket{D}$) and anti-diagonal ($\ket{A}$) polarization.

Within this encoding, the path degree of freedom coherently controls the order in which \AB's operations are applied to the polarization qubit. The two interferometric paths of the quantum SWITCH therefore realize the alternative causal orders, while the polarization carries the quantum information from which the key is extracted. This architecture follows the standard photonic realization of the quantum SWITCH employed in previous experiments and provides a natural platform for implementing the protocol described in Section~\ref{sec:theoretical_framework}. We emphasize here that although the theoretical proposal assumed the control qubit to be private and local, our implementation of the quantum SWITCH realizes neither of these properties. As such, an eavesdropper could access the control qubit, and our implementation is not secure.

Additionally, for this proof-of-principle implementation, we assume that \AB\ can prepare (near-)ideal qubits, thereby neglecting vacuum and multi-photon contributions in each bosonic mode. In practical QKD implementations using weak coherent states, multi-photon emissions can open the door to photon-number-splitting attacks. Countermeasures against such attacks typically require the use of decoy states, which are not considered in the present implementation.

\subsection{Path-Coherence Preserving Measurement Inside the Quantum SWITCH}
\label{subsec:pcp}
Since the shared key bit is encoded in the polarization of the system photon, \A\ can prepare the required states using standard polarization optics. \B, however, must be able to extract the polarization outcome while the photon remains inside the quantum SWITCH. A conventional polarization measurement, for example via a polarizer as in standard optical implementations of BB84, cannot be used. While it can project the photon onto a polarization eigenstate, it does not provide \B\ with a local record of the shared key. In particular, if \B's polarizer is orthogonal to \A's polarizer,
the key is not transmitted.  The challenge is therefore to obtain a classical record of the measurement outcome without disturbing the coherence responsible for the ICO or destroying the photon. To achieve this, we employ the measurement scheme introduced in Ref.~\cite{valibouse2026time}.

The scheme realizes a von Neumann measurement~\cite{von2018mathematical} by coupling the polarization of the system photon to an ancilla photon in \B's possession. Consider a system photon prepared in the normalized state
\begin{equation}
    \ket{\psi}_s = \alpha\ket{H}_s + \beta\ket{V}_s,
\end{equation}
and an ancilla photon prepared in
\begin{equation}
    \ket{+}_a = \tfrac{1}{\sqrt{2}}\big(\ket{H}_a + \ket{V}_a\big).
\end{equation}
Suppose these two photons are injected into the two input ports of a polarizing beam splitter (PBS). Post-selecting for events in which one photon exits in each output port yields
\begin{equation}
    \ket{\psi}_s \otimes\ket{+}_a \;\mapsto\; \ket{\psi'}_{sa} = \alpha\ket{H}_s\ket{H}_a + \beta \ket{V}_s\ket{V}_a.
\end{equation}
The PBS interaction therefore correlates the polarization of the system photon with that of the ancilla, such that the measurement outcome is encoded in the ancilla state. Equivalently, for the ancilla input state $\ket{+}_a$, the post-selected transformation reproduces the action of a CNOT gate, with the system photon acting as the control qubit and the ancilla photon as the target. A subsequent measurement of the ancilla thus realizes a von Neumann measurement of the system polarization. Measurements in the Hadamard basis are implemented by placing half-wave plates oriented at $22.5^\circ$ before and after the PBS on the system path.

Experimentally, implementing this measurement inside a quantum SWITCH presents additional challenges. Performing the system-ancilla coupling independently in each interferometric branch would leave information about the causal-order branch encoded in the ancilla path degree of freedom. Tracing over this information would destroy the coherence between the two alternative orders and hence the ICO itself.
Moreover, when two post-selected gates are applied sequentially, the heralding signal does not distinguish between two successful gates and a failed first gate followed by a successful second detection event. Indeed, a failure of the first gate, in which both photons exit through the same PBS port, can still lead to a coincidence after the second gate and be incorrectly identified as a successful two-gate operation~\cite{barz2014two}.

To avoid this problem, the system and ancilla photons are first entangled in their path degree of freedom. This ensures that, in each branch of the quantum SWITCH, the ancilla photon arrives at the appropriate measurement station and interacts with the system photon at the correct PBS. After the measurement interaction, the ancilla paths are recombined interferometrically, thereby erasing the which-order information carried by the ancilla and restoring coherence between the two causal-order branches.

The required path entanglement is generated using a polarization-entangled photon pair produced by spontaneous parametric down-conversion (SPDC). Each photon is subsequently sent to a PBS, and the polarization is set to a common state for each path so that the remaining entanglement resides solely in the path degree of freedom (see {Figure}~\ref{fig:setup}).

To illustrate the operation of the measurement scheme, consider the joint state of the system and ancilla photons immediately before the measurement interaction. In the most general case it can be written as
\begin{align}
\ket{\psi} &= \frac{1}{2} \Big[
    (\alpha \ket{H,\circlearrowright}_s +\beta\ket{V,\circlearrowright}_s) \otimes (\ket{H,1}_a + \ket{V,1}_a) \notag \\
  &\quad + (\alpha' \ket{H,\circlearrowleft}_s + \beta' \ket{V,\circlearrowleft}_s) \otimes (\ket{H,2}_a + \ket{V,2}_a)
\Big],
\end{align}

where $\circlearrowright$ and $\circlearrowleft$ denote the two causal-order branches of the quantum SWITCH, $1$ and $2$ label the corresponding ancilla paths, and $\alpha', \beta' \in \mathbb{C}$ such that $|\alpha'|^2 + |\beta'|^2 = 1$.

Applying the PBS-based measurement interaction in each branch, followed by post-selection and coherent recombination of the ancilla paths, yields
\begin{align}
    \ket{\psi'} &= \frac{1}{2\sqrt{2}} \Bigl[\alpha\ket{H,\circlearrowright}_s \bigl(\ket{H,1}_a + \ket{H,2}_a \bigr) \notag \\
    &\qquad + \alpha' \ket{H,\circlearrowleft}_s \bigl(\ket{H,1}_a - \ket{H,2}_a \bigr)    \notag \\
    &\qquad +\beta\ket{V,\circlearrowright}_s \bigl(\ket{V,1}_a + \ket{V,2}_a \bigr) \notag \\
    &\qquad + \beta' \ket{V,\circlearrowleft}_s \bigl(\ket{V,1}_a - \ket{V,2}_a \bigr) \Bigr].
\end{align}
The measurement outcome is now encoded in the polarization of the ancilla photon, while coherence between the two causal-order branches is preserved. In this sense, the system polarization can be measured locally without destroying the ICO.
This von Neumann-type measurement establishes correlations between the polarization measurement performed on \B's ancilla photon and the polarization state of \A's system photon, enabling the generation of a shared key. These correlations are preserved for arbitrary measurement bases, as demonstrated in the next section.

After the ancilla-path recombination and the final beam splitter of the quantum SWITCH, the information originally carried by the control qubit is no longer encoded solely in the path degree of freedom of the system photon. Instead, it is distributed across the joint path state of the system and ancilla photons. The logical control states are therefore represented by Bell states in the two-photon path subspace,
\begin{equation}\label{eq:controlBell}
\ket{\Phi^{+}}
=
\frac{\ket{1,1}+\ket{2,2}}{\sqrt{2}},
\qquad
\ket{\Psi^{+}}
=
\frac{\ket{1,2}+\ket{2,1}}{\sqrt{2}},
\end{equation}
which correspond to the logical control states $\ket{+}_c$ and $\ket{-}_c$, respectively. Consequently, measuring the control qubit requires a joint (albeit separable) measurement on the path degrees of freedom of both photons rather than a measurement on a single photonic qubit.

\begin{figure}[t]
    \subfloat[]{\includegraphics[width=.8\linewidth]{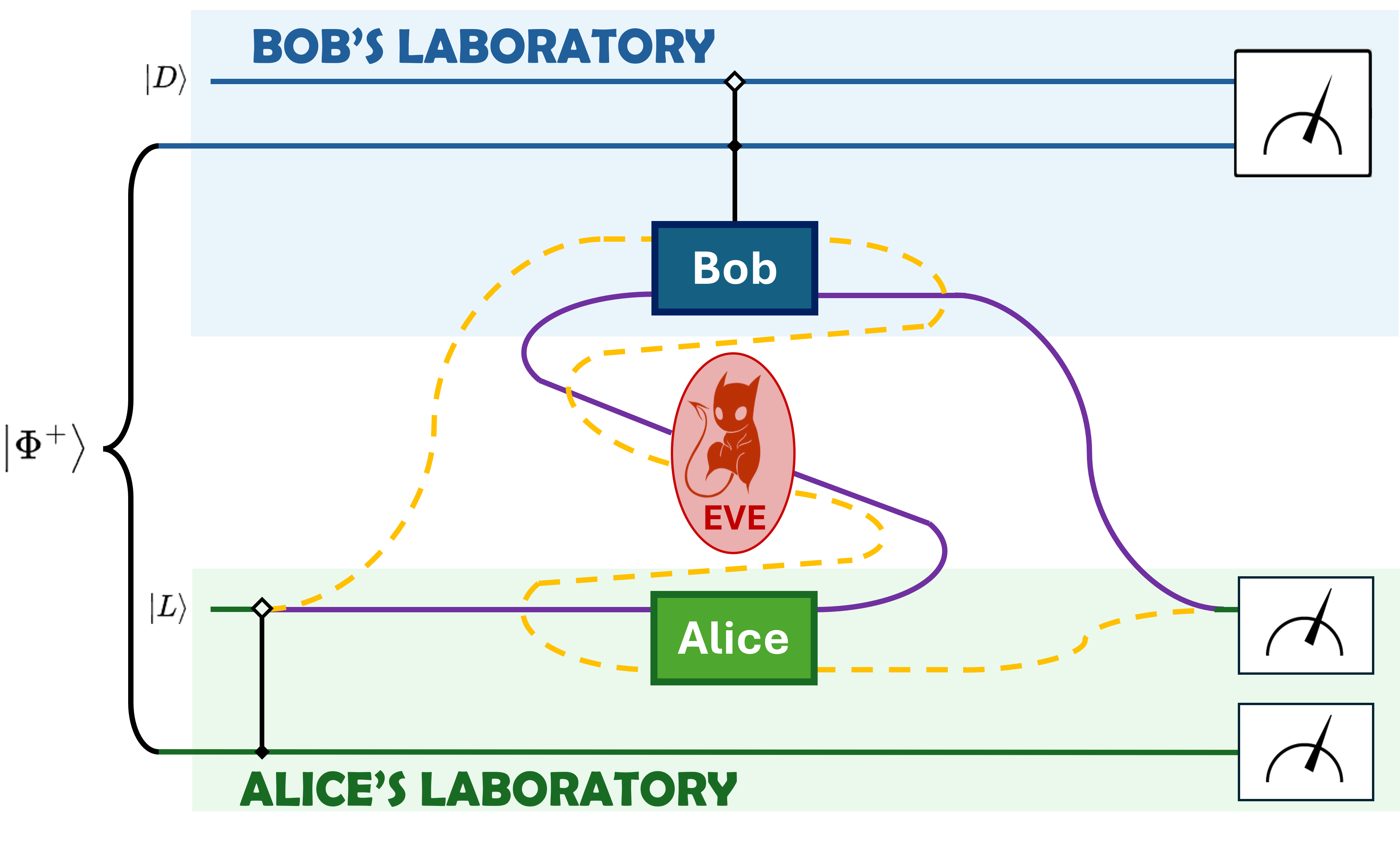}
        \label{fig:a}
    }

    \subfloat[]{\includegraphics[width=.8\linewidth]{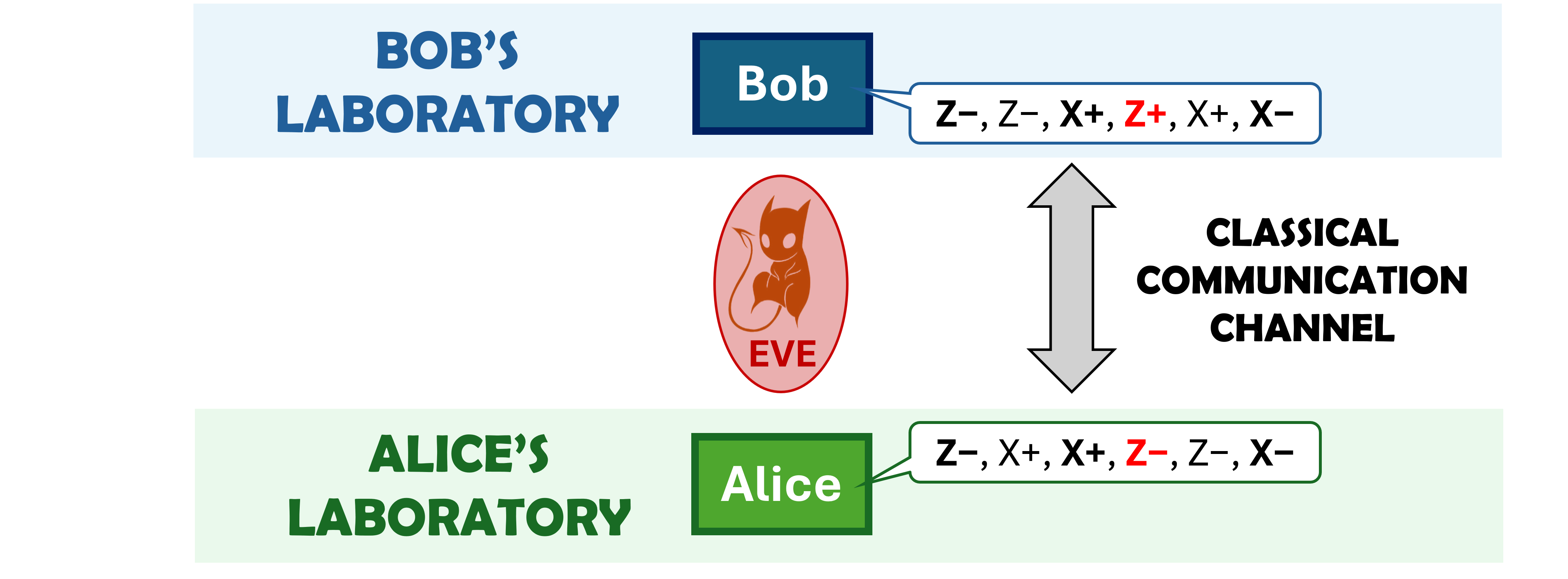}
        \label{fig:b}
    }

    \caption{(\textbf{a}) \AB\ {independently}
 choose a measurement basis and obtain two binary outcomes from each successful event: a polarization outcome, which contributes to the raw key, and a control outcome associated with the output port of their respective interferometers. The latter encodes information about the coherence between the alternative causal-order branches of the quantum SWITCH.
    {(\textbf{b})}
 During the reconciliation stage, \AB\ publicly disclose their basis choices and control outcomes, while keeping the polarization outcomes private. Key bits are retained only when both parties selected the same basis (highlighted in bold). In the absence of an eavesdropper, the control outcomes exhibit definite correlations. Events in which the parties use the same basis but obtain incompatible control outcomes (highlighted in red) indicate a loss of coherence between the causal-order branches and therefore signal the presence of an eavesdropper.
    }
    \label{fig:protocol}
\end{figure}

\subsection{Setup for QKD}

\begin{figure*}[t]
\includegraphics[width=\linewidth]{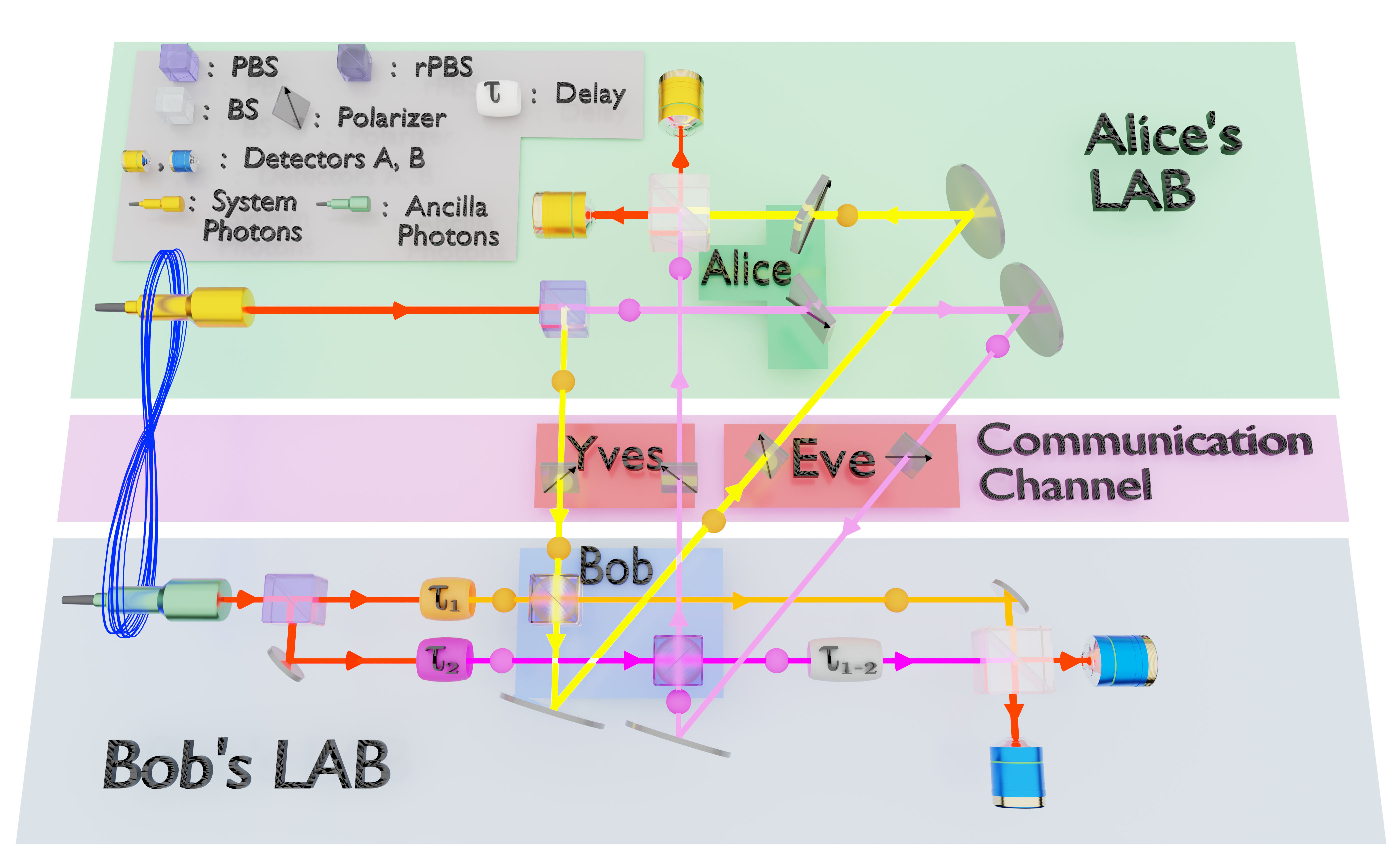}
\caption{{Experimental}
 realization of the quantum SWITCH with a path-coherence-preserving measurement performed in \B's laboratory. The yellow and blue detectors correspond to measurements performed by \A\ and \B, respectively. The protocol admits two distinct locations at which an eavesdropper may intercept the communication, denoted by \E\ and \Y, corresponding to the two possible insertion points within the alternative causal-order branches. In the present experiment, only the eavesdropping strategy associated with \E\ is implemented. On \A's side, the encoding of the key happens in the darker green area with polarizers while for \B, the encoding happens via rotated PBSs (rPBS: PBSs surrounded by half-wave plates) in the darker blue area. The source of photons is placed in \A's lab. The pink and yellow paths inside the quantum SWITCH correspond to the two superposed causally ordered branches and are associated with the orange and fuchsia paths of the ancilla, respectively. The time delays $\tau_1$ and $\tau_2$ are chosen such that the ancilla photon arrives at the PBSs implementing the entangling gate simultaneously with the system photon. The delay $\tau_{1-2}$, however, is part of the erasure protocol and is chosen such that no which-branch information is revealed upon measuring the ancilla.}
\label{fig:setup}
\end{figure*}

The experimental setup is shown in Figure~\ref{fig:setup}. Once inside the quantum SWITCH (i.e.,\,beyond the PBS in \A's lab), \A\ prepares the system polarization in one of the four possible BB84 states \eqref{eq:statesBB84} (using polarization optics only) and stores the corresponding bit value $b$ and basis choice $\mu$
in a classical register. She also records the outcome of the control qubit measurement from the detector that clicks at the output of the quantum SWITCH. The input polarization of the photon is prepared in the left-circular state
\begin{equation*}
    \ket{L} = \frac{1}{\sqrt{2}}\left(\ket{H} + i\ket{V}\right),
\end{equation*}
so that the probability of transmission through \A's polarizers is independent of whether they are oriented in the computational or Hadamard basis.

A central requirement of the protocol is that \B\ must also obtain the value of the shared key bit, despite having access to neither the input nor the output of the quantum SWITCH. This is achieved using the path-coherence-preserving measurement protocol described in Section~\ref{subsec:pcp}. The ancilla photon required for this measurement is generated together with the system photon in the SPDC source and routed to \B's laboratory. There, it interacts with the system photon through the post-selected measurement gate, allowing \B\ to extract the polarization outcome while preserving the coherence between the alternative causal-order branches. Consequently, each successful event provides both \AB\ with access to the same key bit.

It is important to emphasize that the ancilla photon does not carry any information about \A's encoded state. Before the measurement interaction, the ancilla polarization is always prepared in the fixed state $\ket{D}$ and therefore contains no information about the key. The key information is transferred to the ancilla only through its interaction with the system photon during \B's measurement procedure. The entanglement generated by the SPDC source is nevertheless essential: although the ancilla polarization is fixed, the system and ancilla photons are entangled in their path degree of freedom. This path entanglement allows the ancilla to be coherently associated with the two causal-order branches and enables the subsequent erasure of which-order information. Thus, the initial entanglement is not a source of key information, but a resource required to implement the path-coherence-preserving measurement when post-selected gates are used.

Because the present implementation relies on pre-entanglement and post-selected linear-optical gates, it should not be regarded as a realization of a QKD protocol with proven security. Rather, it constitutes a proof-of-principle experimental demonstration of how ICO can be exploited for QKD.

In particular, the entangling operations are implemented using probabilistic linear-optical gates, since linear optics alone cannot realize deterministic CNOT gates. Consequently, the protocol relies on post-selection, leaving the post-selection security loophole open. Closing this loophole would require replacing the probabilistic linear-optical entangling gates with deterministic photon--photon interactions enabled by nonlinear light--matter interfaces, such as cavity QED or other matter-mediated interaction schemes~\cite{duan2004scalable,hacker2016photon}. A more detailed discussion of these aspects can be found in Ref.~\cite{valibouse2026time}.

In addition to the polarization bit used for key generation, \A\ and \B\ each obtain a second binary outcome associated with the output port of their respective interferometers, as shown in Figure~\ref{fig:protocol}a. These outcomes encode the information carried by the control degree of freedom and can be publicly communicated together with the basis choices during the reconciliation stage, as illustrated in Figure~\ref{fig:protocol}b. After \B's measurement, the control information is encoded in the Bell states $\ket{\Phi^+}$ and $\ket{\Psi^+}$ of the joint path degree of freedom defined in Equation~\eqref{eq:controlBell}.
To distinguish these two states, \AB\ measure in the computational basis, detecting either correlated ($\ket{\Phi^+}$) or anti-correlated ($\ket{\Psi^+}$) outcomes.
These outcomes characterize the relative phase between the two causal-order branches, but are independent of the polarization outcomes that form the raw key. In particular, distinguishing the states $\ket{\Phi^+}$ and $\ket{\Psi^+}$ does not reveal whether the polarization outcome was $H$, $V$, $D$, or $A$. Therefore, the control outcomes can be publicly disclosed without revealing key information, while still providing a signature of eavesdropping.

The role of the control outcomes can be seen from the ideal shared state in the absence of an eavesdropper. When \AB\ use a common basis, and \A\ prepares the polarization state $P$, with $P\in\{H,V,D,A\}$, the shared two-photon state takes the form
\begin{equation}\label{eq:sharedSt}
    \ket{\psi}_{\text{shared}} = \frac{1}{\sqrt{2}}(\ket{P,+}_A\ket{P,+}_B + \ket{P,-}_A\ket{P,-}_B),
\end{equation}
where $\ket{\pm}_{A,B}$ denote the two output-port outcomes associated with the control measurement on \A's and \B's sides. This state guarantees both agreement of the polarization key bit and perfect correlations between the control outcomes. A disturbance that reduces the interference between the causal-order branches can populate output-port combinations that are forbidden in the honest case, thereby revealing the presence of an eavesdropper. A derivation of the corresponding joint state, including the classical registers, polarization degrees of freedom, and control degrees of freedom, is given in Appendix~\ref{app:sharedstate}.

We implement one of the eavesdroppers (\E\ in Figure~\ref{fig:setup}) by placing identical polarizers at an angle $\theta$ (and $\theta + 90^{\circ}$ for the corresponding orthogonal outcome) in both paths of the quantum SWITCH. This simulates an intercept-resend-type attack where \E\ performs a controllable projective measurement on the polarization degree of freedom while being prevented from directly accessing the control system. By varying $\theta$, we change the basis in which \E\ probes the photon and therefore the disturbance introduced into the shared state \eqref{eq:sharedSt}, and by repeating at an angle $\theta + 90^{\circ}$, the statistics of both outcomes of a projective measurement in such a basis can be simulated. In principle, \E's projective measurement could be truly realized using the same path-coherence-preserving measurement protocol described in Section~\ref{subsec:pcp} as \B. Such an implementation, however, would require an additional ancilla system and is substantially more experimentally demanding. Note that our experimental setup focuses on \E\ and does not implement \Y, although he would be detected via the same protocol as for \E. Furthermore, the theoretical proposal from Ref.~\cite{spencer2025indefinite} assumes that the control qubit is private and stays in \A's laboratory. To date, no such implementation of the quantum SWITCH has been realized. Therefore, we only simulate \E's ignorance of the control-qubit state by making her act in the same way, independently of the causal-order branch.

\section{Experimental Results}
\label{sec:results}

\subsection{Without Eavesdropper}
\label{subsec:withoutEve}

\begin{figure}[t]
    \includegraphics[width=.95\linewidth]{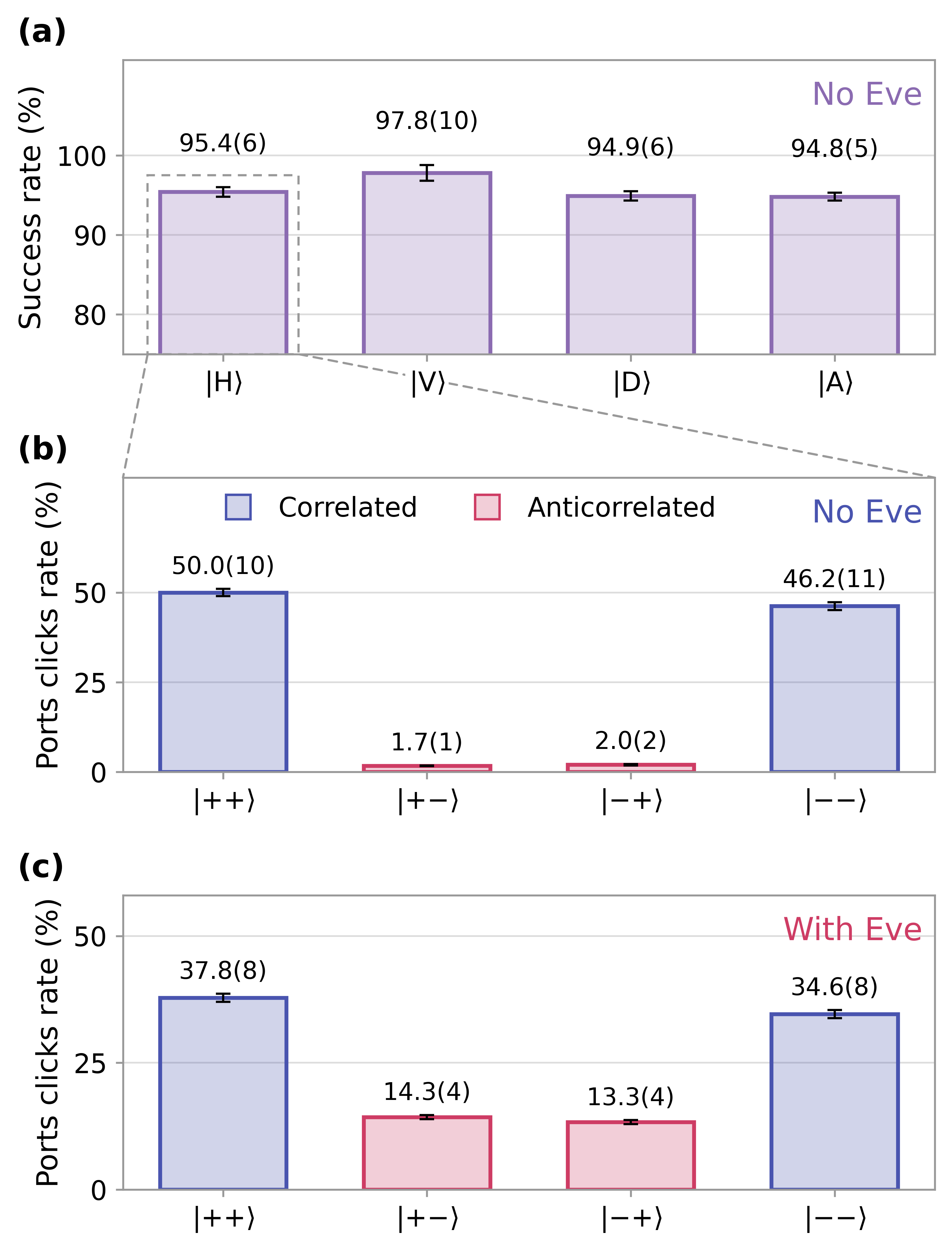}
    \caption{{Measured}
 protocol performance, without and with an eavesdropper.
    (\textbf{a}) Probability that \AB\ obtain the same polarization outcome, and therefore establish a shared key bit, for the four BB84 signal states. The purple bars show this success rate in the absence of \E.
    (\textbf{b}) Port-resolved breakdown of the $\ket{H}$ case highlighted in (\textbf{a}), indicated by the dashed guides. Blue bars denote correlated control outcomes, which yield key material; red bars denote anti-correlated control outcomes, which correspond to false-positive eavesdropping detections and would ideally vanish. They provide a measure of the intrinsic false-positive rate of the implementation.
    (\textbf{c}) The same port-resolved measurement in the presence of \E. Here, \A\ prepares the state $\ket{H}$ and \E's polarizers are set at $\theta = 45^\circ$, corresponding to a measurement in the diagonal basis. The theoretical detection probability is $p_{\mathrm{detect}|0,0}(45^\circ) = 25\%$, while the experimentally observed value is $(27.6 \pm 1.4)\%$.}
    \label{fig:ports_rate}
\end{figure}

We first characterize the performance of the protocol in the absence of an eavesdropper. Figure~\ref{fig:ports_rate}a reports the measured probability that \AB\ obtain the same key bit after basis reconciliation for the four BB84 signal states $\ket{H}$, $\ket{V}$, $\ket{D}$, and $\ket{A}$. Figure~\ref{fig:ports_rate}b resolves the individual output-port combinations for the state $\ket{H}$; the anti-correlated control outcomes should ideally vanish and therefore provide a measure of the intrinsic false-positive eavesdropping-detection rate of the implementation.

In the absence of an eavesdropper, \AB\ successfully establish a shared key bit with probability $0.964 \pm 0.004$. Under the same conditions, the false-positive eavesdropper-detection probability is $0.033 \pm 0.002$. The false-positive rate arises from imperfect visibility of the quantum SWITCH.

\subsection{With an Eavesdropper}

We now investigate the performance of the protocol in the presence of an eavesdropper. As discussed in Section~\ref{sec:theoretical_framework}, \E's intervention reduces the coherence between the two causal-order branches of the quantum SWITCH. The corresponding joint state shared by \A, \B, and \E\ is derived explicitly in Appendix~\ref{app:sharedstate}. Experimentally, this manifests itself as an increased probability of observing the anti-correlated control outcomes, which therefore serves as the signature of eavesdropping.

To describe the detection probability, let $b\in\{0,1\}$ denote the bit encoded by \A\ and recall that $\mu\in\{0,1\}$ denotes the common basis (computational or Hadamard) chosen by \AB\ after reconciliation. We define:
\begin{equation}\label{eq:thetaMu}
    \theta_{b,\mu} = \theta + \varphi_b + \phi_\mu,
\end{equation}
where
\begin{equation}
    \varphi_0 = 0^\circ, \qquad \varphi_1 = 90^{\circ},
\end{equation}
and
\begin{equation}
    \phi_0 = 0^\circ, \qquad \phi_1 = -45^{\circ},
\end{equation}
where, as described earlier, $\theta$ is the orientation of \E's projective measurement.

As derived in Appendix~\ref{app:detProb}, the probability of detecting \E\ is
\begin{equation}
    p_{\mathrm{detect} | b,\mu}(\theta) =
    \cos^2(\theta_{b,\mu})\sin^2(\theta_{b,\mu}).
\end{equation}
A representative example is shown in Figure~\ref{fig:ports_rate}c, where \A\ prepares the state $\ket{H}$, corresponding to $b = 0$ and $\mu = 0$, and \E\ measures in the diagonal basis, corresponding to $\theta = 45^\circ$. In this configuration, the theory predicts
\begin{equation}\label{eq:detectH}
    p_{\mathrm{detect} | 0, 0}\left(45^\circ\right) =
    \frac{1}{4},
\end{equation}
because \E's basis is mutually unbiased with respect to the state prepared by \A. Experimentally, we obtain
\begin{equation}
    p_{\mathrm{detect}| 0,0} \left( 45^\circ \right) = 0.276 \pm 0.014,
\end{equation}
in good agreement with \eqref{eq:detectH}.

\begin{figure}[t]
    \includegraphics[width=.95\linewidth]{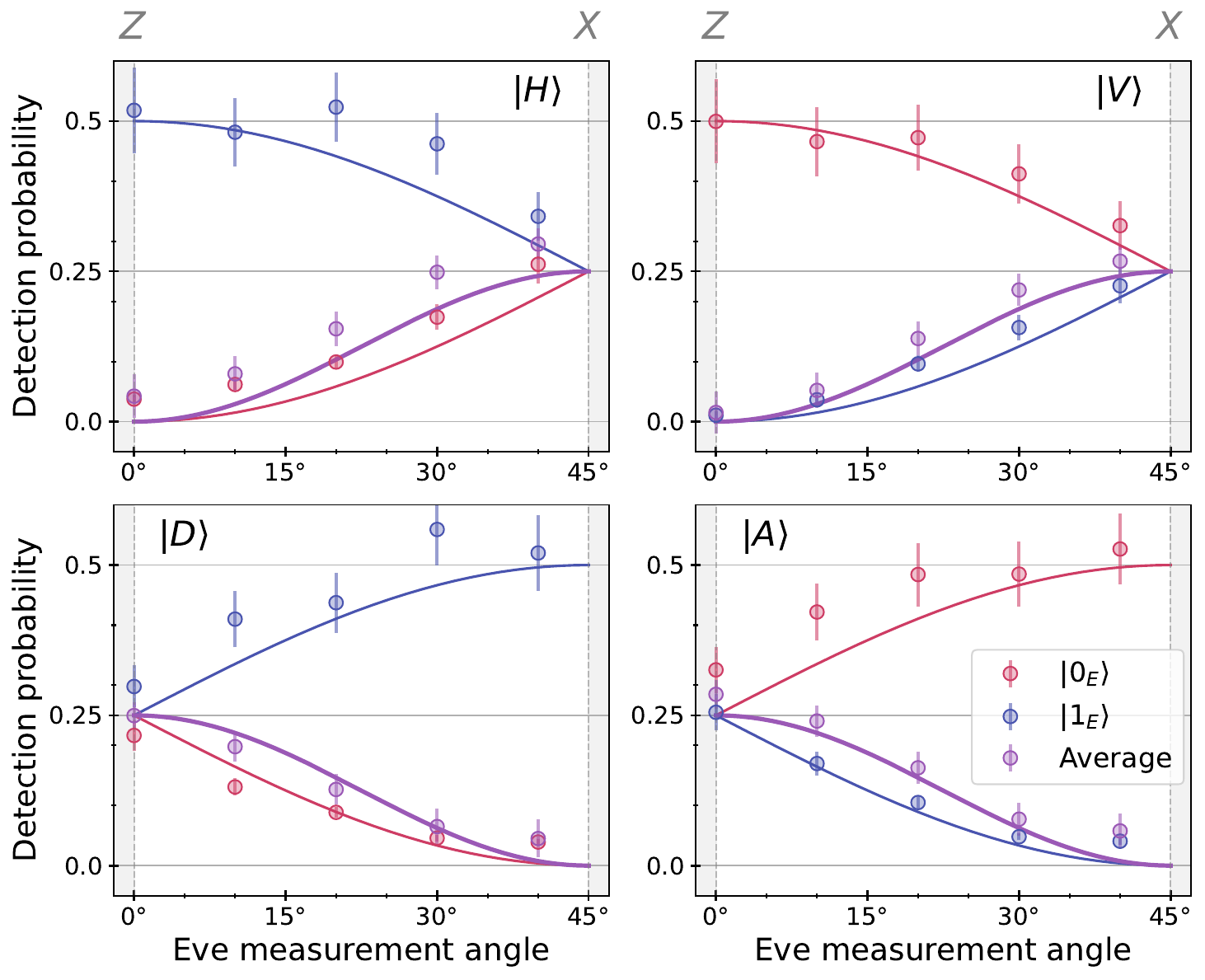}
    \caption{Detection probability as a function of \E's polarizer angle $\theta$ for the four BB84 signal states \eqref{eq:statesBB84}. Red and blue data correspond to the conditional detection probabilities $\mathbb{P}_{\mathrm{detect}|0,b,\mu}$ and $\mathbb{P}_{\mathrm{detect}|1,b,\mu}$,
    associated with \E's two possible measurement outcomes. Purple data show the detection probability obtained after averaging over \E's outcomes. Solid curves are the theoretical predictions.}
    \label{fig:p_detec}
\end{figure}

Since \E\ has no prior information about which of the four BB84 signal states \eqref{eq:statesBB84} was prepared by \A, we average the detection probability over all possible basis choices and bit values. This yields the theoretical average detection probability
\begin{equation}\label{eq:detectAvg}
    p_{\mathrm{detect}} = \frac{1}{8},
\end{equation}
as shown in Appendix~\ref{app:detProb}. The complete set of measured detection probabilities is shown in Figure~\ref{fig:p_detec} for all four BB84 signal states and for both possible outcomes of \E's measurement.
Averaging over all four BB84 states yields
\begin{equation}
    p_{\mathrm{detect}} = 0.15 \pm 0.02,
\end{equation}
which is slightly larger than the ideal value \eqref{eq:detectAvg}. This discrepancy is consistent with the false-positive detection probability reported in Section~\ref{subsec:withoutEve} and can be attributed primarily to the finite interference visibility of the quantum SWITCH.

Experimentally, \E\ is implemented with polarizers. A single polarizer setting accesses only one of the two possible outcomes of \E's measurement, namely the polarization component transmitted by the polarizer. Consequently, a single experimental run yields only the contribution associated with one value of \E's classical register. To emulate a complete intercept-resend attack, we repeat the experiment with \E's polarizers rotated by $90^\circ$, thereby obtaining the complementary outcome. The two data sets are then combined to reconstruct the statistics of an eavesdropper capable of both measuring and re-preparing the quantum state.

\subsection{Mutual Information}
\label{subsec:mutInfo}

To quantify the correlations between the parties, we compute the mutual information between their classical registers. This quantity measures how much information one party can infer about another party's key bit. We evaluate both the mutual information shared by \A\ and \B, and the information acquired by \E\ through her measurement.

\begin{figure}[t]
    \includegraphics[width=.95\linewidth]{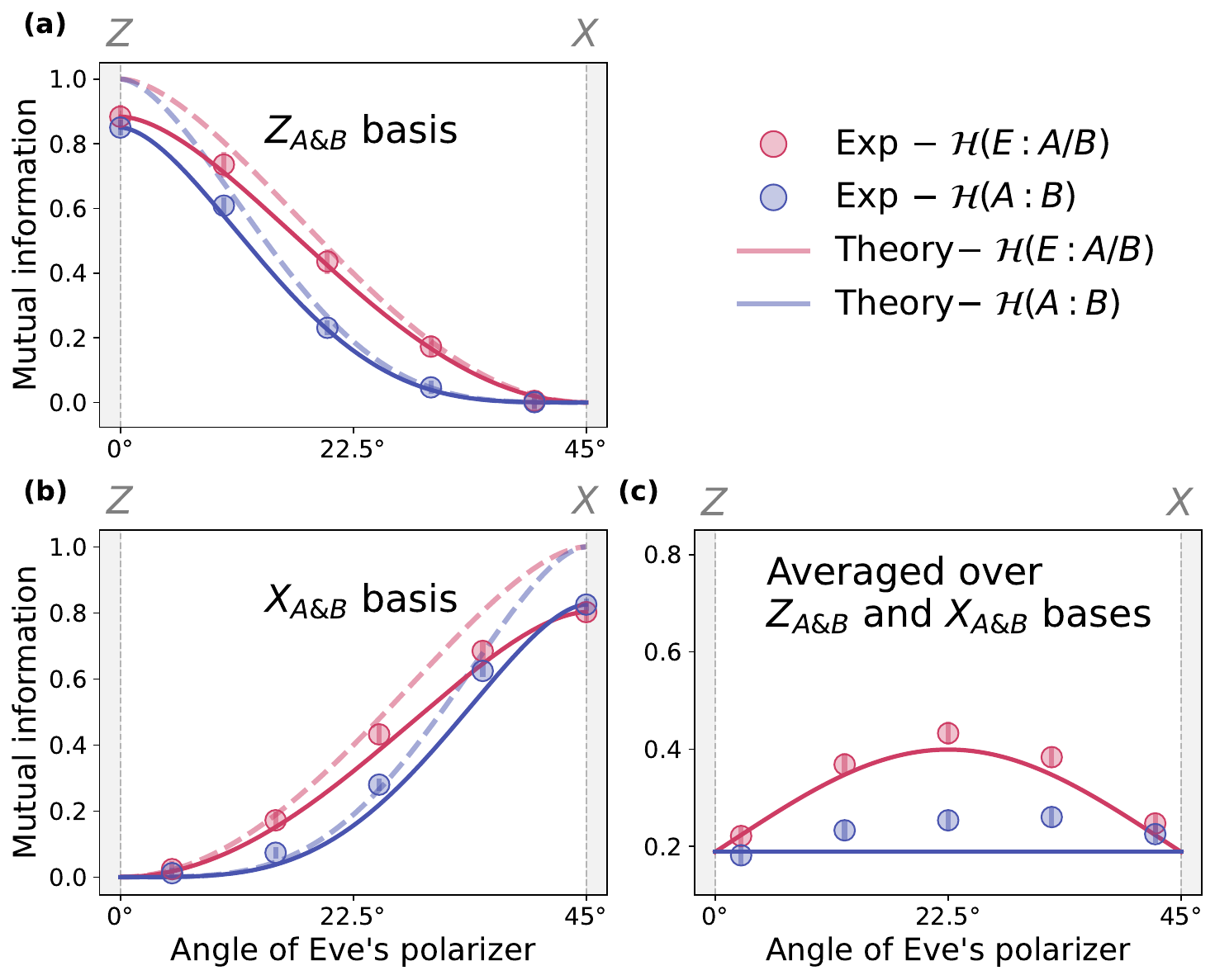}
    \caption{(\textbf{a,b}) Mutual information between the different parties when \A\ and \B\ use (\textbf{a}) the computational basis and (\textbf{b}) the Hadamard basis. Dashed curves correspond to the theoretical predictions, while solid curves show the same predictions renormalized using the experimentally observed maximum mutual information.
    (\textbf{c}) Mutual information averaged over all four BB84 signal states \eqref{eq:statesBB84} as a function of \E's polarizer angle $\theta$. The mutual information between \A\ and \B\ is constant and equal to $\mathcal{H}(A:B) = -1 + \tfrac{3}{4}\log_2 3\simeq 0.1887$, whereas \E's mutual information with \A\ or \B\ is maximal for $\theta = 22.5^\circ$.}
    \label{fig:mutual}
\end{figure}

For two parties $X$ and $Y$, with $X, Y \in \{\A,\;\B,\;\E\}$, the mutual information is
\begin{align}\label{eq:mutualInfo}
    \mathcal{H}(X:Y) =
        &-\sum_{i\in \{0,1\}}\mathbb{P}(x_i)\log_2\mathbb{P}(x_i) \notag \\
        &-\sum_{j\in \{0,1\}}\mathbb{P}(y_j)\log_2\mathbb{P}(y_j) \notag \\
        &+\sum_{i,j\in \{0,1\}}\mathbb{P}(x_i, y_j)\log_2\mathbb{P}(x_i,y_j),
\end{align}
where $\mathbb{P}(\cdot)$ denotes the relevant marginal or joint probability distribution. In our experiment, the classical registers are binary and their marginal distributions are uniform. The mutual information therefore reduces to
\begin{align}
    \nonumber \mathcal{H}(X:Y) &= 1 + p_{X,Y}\log_2(p_{X,Y}) \\
    \qquad &+ (1-p_{X,Y})\log_2(1-p_{X,Y}),
\end{align}
where $p_{X,Y}$ is the probability that the two parties obtain the same bit value (see Appendix~\ref{app:mutuinfo}). For example, when \AB\ use the computational basis, the bit agreement probability is
\begin{equation}
    p_{A,B}(\theta) = 1 - 2\cos^2\theta\sin^2\theta,
\end{equation}
while \E's bit agreement probability with respect to \A\ or \B\ is
\begin{equation}
    p_{E,A/B}(\theta) = \cos^2\theta.
\end{equation}
The resulting mutual informations are shown in Figure~\ref{fig:mutual}a,b for the two measurement bases.

To obtain an overall characterization of the protocol, we average the mutual information over the four BB84 signal states \eqref{eq:statesBB84} prepared by \A. Equivalently, we assume that \E\ attacks every transmitted photon but has no prior information about either the basis or the bit chosen by \A. The resulting mutual informations are shown in Figure~\ref{fig:mutual}c. Under this averaging, the mutual information shared by \AB\ becomes independent of \E's polarizer angle $\theta$. In contrast, the mutual information between \E\ and either legitimate party depends on $\theta$ and reaches a maximum at $\theta = 22.5^\circ$. This angle therefore corresponds to \E's optimal measurement setting, maximizing the information she can extract on average from the key.

The previous analysis assumes that \E\ intercepts every photon transmitted through the protocol. A more subtle scenario is one in which \E\ attacks only a fraction of the signals. To model this situation, we introduce the parameter $t \in [0,1]$ that quantifies the strength of the eavesdropping attack. The limiting cases are $t = 0$, corresponding to the absence of an eavesdropper, and $t = 1$, corresponding to \E\ measuring every photon. Operationally, this can be interpreted as \E\ inserting her polarizers for a fraction $t$ of the runs and removing them for the remaining fraction $1-t$.

\begin{figure}[t]
    \includegraphics[width=.95\linewidth]{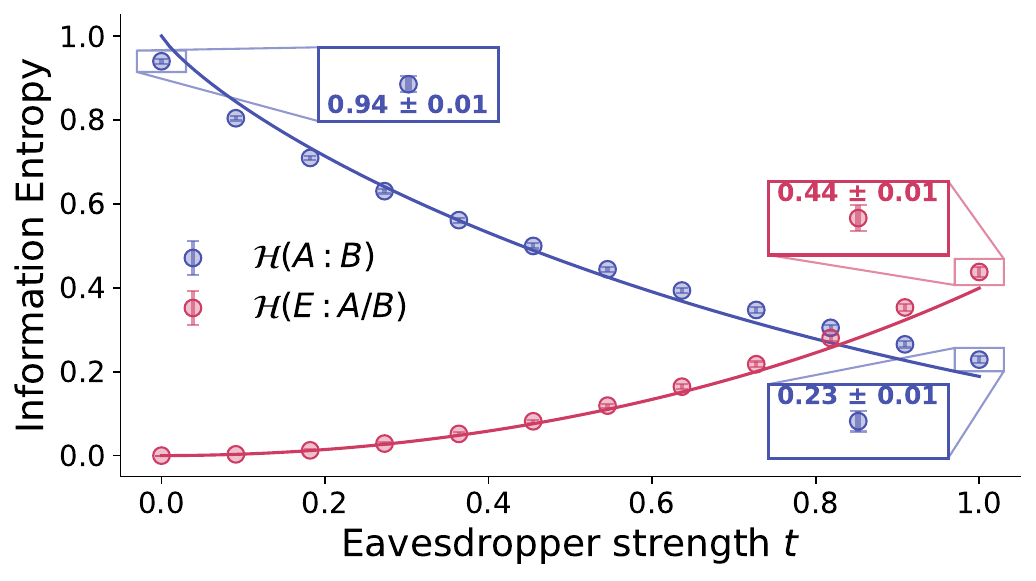}
    \caption{Mutual information as a function of the eavesdropping strength $t$ for \E's optimal fixed measurement strategy, corresponding to a polarizer angle $\theta = 22.5^\circ$. At $t\simeq 0.8284$, corresponding to a detection probability $p_{\mathrm{detect}}\simeq 0.1036$, \E's mutual information about the key exceeds the mutual information shared by \AB. Insets show the experimentally measured data points; the remaining points are obtained from the interpolation model described in the text.}
    \label{fig:mutu_opti}
\end{figure}

Since we can experimentally determine the mutual information between \AB\ when \E\ is absent, as well as the mutual information between all parties when \E\ performs her optimal measurement, we can construct the mutual-information curves for any eavesdropping strength $t$. The effective agreement probabilities used to compute the mutual information are
\begin{equation}\label{eq:strengthEve}
    p_{X,Y}(t) = (1-t)\,p^{(\text{no Eve})}_{X,Y} \;+\; t\,p^{(\text{with Eve})}_{X,Y}.
\end{equation}
The corresponding mutual informations are shown in Figure~\ref{fig:mutu_opti} for \E's optimal measurement angle, $\theta = 22.5^\circ$. We find a threshold value
\begin{equation}
    t_0 \simeq 0.8284,
\end{equation}
above which \E\ obtains more information about the key than \AB\ share with each other. This corresponds to a detection probability
\begin{equation}
    p_{\mathrm{detect}}\simeq 0.1036.
\end{equation}

\section{Conclusions}
\label{sec:conclusions}

We have reported the first experimental proof-of-principle demonstration of a quantum key distribution protocol in an indefinite causal order. By embedding the communicating parties, \AB, within a photonic quantum SWITCH, we implemented a BB84-inspired protocol in which the presence of an eavesdropper is detected through measurements of the control qubit rather than through the public comparison of the raw key. In the absence of an eavesdropper, the protocol\textemdash upon successful post-selection\textemdash achieved a key-generation success probability of $(96.4 \pm 0.4)\%$. When an eavesdropper, implementing a coherent measure-and-reprepare attack, was introduced inside the quantum SWITCH, we measured an average detection probability of $0.15 \pm 0.02$ per qubit, in good agreement with the theoretical prediction of 0.125.

Unlike the standard BB84 scheme, eavesdropper detection in the present protocol does not require publicly revealing and discarding a subset of the raw key. Instead, the information used to detect an attack is carried entirely by the control degree of freedom, while the polarization outcomes remain available for key generation. \AB\ still perform the usual classical reconciliation of their basis choices and control outcomes, but this communication reveals no information about the shared key.

Although our present work does not derive a secure key rate\textemdash which would require the categorization and optimization of all possible attacks while also considering experimental imperfections\textemdash we have shown that specific eavesdropper attacks can be detected by honest parties with a non-negligible probability.
Crucially, this cheat sensitivity is at the core of the general security advantage provided by quantum cryptography, and therefore already elegantly demonstrates how ICO may be used to construct a QKD protocol with some fundamental passive security advantage over its classical counterparts.

The present implementation relies on post-selected linear-optical operations, together with pre-entanglement between the system and ancilla photons, which can open side channels for a variety of new attacks; therefore, further work is required for the full verification of security.
Additionally, one key feature of all photonic quantum SWITCH implementations to date is that the control qubit is encoded in the same photon as the system qubit. This opens an attack where \E\ can learn about the control qubit by a suitable measurement of the path and can adapt her actions accordingly.
Keeping the control qubit private is a long-standing goal of experimental studies of ICO. Some suggest that the use of single-photon-level light--matter interactions could allow one to achieve this~\cite{rozema_review}. Such interactions are possible in principle~\cite{duan2004scalable, hacker2016photon}. Our experimental implementation relies on linear optics and therefore cannot hide the state of the control qubit.
Nevertheless, this work establishes the experimental feasibility of implementing quantum cryptographic protocols within an indefinite causal order and demonstrates that it can serve as an alternative resource for eavesdropper detection. We expect that future implementations avoiding post-selection, together with a complete security analysis against general attacks and a rigorous accounting of experimental imperfections such as channel loss, multiphoton emission, detector dark counts, etc., will clarify the practical advantages that indefinite causal order may offer for quantum cryptography.

\section*{Data and Code Availability}
All the data and code that
are necessary to replicate, verify, falsify and/or reuse
this research are available online at \cite{dataset}.

\begin{acknowledgments}
We thank Jessica Bavaresco, Cosmo Lupo, Marco Tulio Quintino, and Teodor Strömberg for insightful discussions.
This project has received funding from the European Union (ERC, GRAVITES, No 101071779), the European Union’s Horizon 2020 research and innovation programme under grant agreement No 899368 (EPIQUS), the European Union’s Horizon 2020 research and innovation programme under the Marie Skłodowska-Curie grant agreement No 956071 (AppQInfo) and the European Union (HORIZON Europe Research and Innovation Programme, EPIQUE, No 101135288). This research was funded in whole or in part by the Austrian Science Fund (FWF) [10.55776/COE1] (Quantum Science Austria), [10.55776/F71] (BeyondC) and [10.55776/FG5] (Research Group 5). This research was funded in whole or in part by the Austrian Science Fund (FWF) 10.55776/PAT4559623. This material is based upon work supported by the Air Force Office of Scientific Research under award numbers FA9550-21-1-0355 (Q-Trust) and FA8655-23-1-7063 (TIQI). The views and opinions expressed are, however, those of the author(s) only and do not necessarily reflect those of the European Union or the European Research Council Executive Agency. Neither the European Union nor the granting authority can be held responsible for them. For open access purposes, the author has applied a CC BY public copyright license to any author accepted manuscript version arising from this submission. The financial support provided by the Austrian Federal Ministry of Labour and Economy, the National Foundation for Research, Technology and Development and the Christian Doppler Research Association is gratefully acknowledged. This work benefitted from network activities through the INAQT network, supported by the Engineering and Physical Sciences Research Council (Grant No. EP/W026910/1). The financial support provided by The Engineering and Physical Sciences Research Council under grant number EP/Y014456/1 is acknowledged.
Open Access Funding by the FWF's Open-Access Block Grant.
\end{acknowledgments}

\newpage

\bibliography{bibliography}

\begin{widetext}

\newpage
\appendix

\section[\appendixname~\thesection]{Shared State Inside the Quantum SWITCH}
\label{app:sharedstate}

In this Appendix, we write the joint state of the relevant classical registers, polarization degrees of freedom, and path degrees of freedom inside the quantum SWITCH. \A\ uniformly chooses one of the four BB84 states \eqref{eq:statesBB84},
\[
    \ket{0}_s\equiv\ket{H}, \qquad \ket{1}_s\equiv\ket{V}, \qquad \ket{+}_s\equiv\ket{D}, \qquad \ket{-}_s\equiv\ket{A}.
\]
Similarly to Section \ref{sec:theoretical_framework}, we denote by $b\in\{0,1\}$ the key bit and $\mu\in\{0,1\}$ the bit specifying the measurement basis, and we denote by $\ket{b,\mu}_{R_A}$ and $\ket{b,\mu}_{R_B}$ the classical registers of \AB. The polarization degrees of freedom associated with \AB's photons are denoted by $Q_A$ and $Q_B$, respectively. \E's classical register is denoted by $R_E$. Finally, the path degrees of freedom associated with the distributed control measurement are denoted by $p_A$ and $p_B$.

The classical registers encode the outcomes of the polarization measurements and therefore the classical information available to the different parties. Throughout this Appendix, we consider the state after \AB\ have publicly compared their basis choices and discarded all events for which different bases were used.

In the absence of an eavesdropper, the shared state can be written as
\begin{align}
    \ket{\Psi}_{\mathrm{shared}} = \frac{1}{2}\Big( &\ket{0,0}_{R_A}\ket{0,0}_{R_B} \ket{H}_{Q_A}\ket{H}_{Q_B} + \ket{1,0}_{R_A}\ket{1,0}_{R_B} \ket{V}_{Q_A}\ket{V}_{Q_B} \nonumber\\
    &+ \ket{0,1}_{R_A}\ket{0,1}_{R_B} \ket{D}_{Q_A}\ket{D}_{Q_B} + \ket{1,1}_{R_A}\ket{1,1}_{R_B} \ket{A}_{Q_A}\ket{A}_{Q_B} \Big) \nonumber \\
    &\otimes \ket{\Phi^+}_{p_Ap_B} \otimes \ket{\mathrm{ignorant}}_{R_E},
\end{align}
where
\begin{equation}
    \ket{\Phi^+}_{p_Ap_B} = \frac{ \ket{0}_{p_A}\ket{0}_{p_B} + \ket{1}_{p_A}\ket{1}_{p_B} }{\sqrt{2}}.
\end{equation}
This expression makes explicit that, in the honest case, \AB\ share identical polarization outcomes, while \E's register is uncorrelated with the key.

\E\ is implemented by polarizers acting identically on the two causal-order branches of the quantum SWITCH. A polarizer oriented at an angle $\theta$ with respect to the horizontal polarization is described by the projector
\begin{equation}
    P_\theta = \ket{\theta}\!\bra{\theta}, \qquad \ket{\theta} = \cos\theta\ket{H} + \sin\theta\ket{V}.
\end{equation}
In the $\{\ket{H},\ket{V}\}$ basis, this projector has the Jones representation
\begin{equation}
    P_\theta = \begin{pmatrix}
        \cos^2\theta & \cos\theta \sin\theta \\
        \cos\theta\sin\theta & \sin^2\theta
    \end{pmatrix}.
\end{equation}
The orthogonal outcome is obtained by rotating the polarizer by $\pi/2$.

Since \E\ is implemented experimentally using passive polarizers, a single polarizer setting gives access only to the transmitted component, which we associate with the register state $\ket{0}_{R_E}$. To reconstruct the complementary component, associated with $\ket{1}_{R_E}$, we repeat the experiment with \E's polarizers rotated by $\pi/2$. Combining the two data sets gives the statistics of an intercept-resend eavesdropper measuring in the basis $\{\ket{\theta}, \ket{\theta+\pi/2}\}$. For compactness, we define
\begin{equation}
    \theta_\times = \theta-\frac{\pi}{4},
\end{equation}
which is \E's polarizer angle expressed relative to the Hadamard basis. The resulting post-selected state in the presence of \E\ can be written as

{
\begin{align}\label{eq:app:sharedStateEve}
\ket{\Psi}_{\mathrm{shared}}^{(E)} &=\frac{1}{2}\ket{0,0}_{R_A}\bigg( \notag \frac{\cos\theta}{2}\ket{0}_{R_E}
    \bigg[
        2\cos\theta \ket{0,0}_{R_B}\ket{H,H}_{Q_A,Q_B}\ket{\Phi^+}
        \\
    &\qquad\qquad + \sin\theta\,\ket{1,0}_{R_B}\ket{H,V}_{Q_A,Q_B} \big(\ket{\Phi^+} + \ket{\Psi^+}\big)
        + \sin\theta\,\ket{1,0}_{R_B}\ket{V,V}_{Q_A,Q_B} \big(\ket{\Phi^+} - \ket{\Psi^+}\big)
    \bigg] \notag\\
\nonumber &\qquad
    + \frac{\sin\theta}{2}\ket{1}_{R_E}
    \bigg[
        2\sin\theta \ket{0,0}_{R_B}\ket{H,H}_{Q_A,Q_B}\ket{\Phi^+}
        \\
    &\qquad\qquad + \cos\theta\,\ket{1,0}_{R_B}\ket{H,V}_{Q_A,Q_B} \big(\ket{\Phi^+} + \ket{\Psi^+}\big)
        + \cos\theta\,\ket{1,0}_{R_B}\ket{V,V}_{Q_A,Q_B} \big(\ket{\Phi^+} - \ket{\Psi^+}\big)
    \bigg]
\bigg) \notag\\
\nonumber &+ \frac{1}{2}\ket{0,1}_{R_A}\bigg( \notag \frac{\cos\theta_\times}{2}\ket{0}_{R_E}
    \bigg[
        2\cos\theta_\times \ket{0,1}_{R_B}\ket{H,H}_{Q_A,Q_B}\ket{\Phi^+}
        \\
    &\qquad\qquad+ \sin\theta_\times\ket{1,1}_{R_B}\ket{H,V}_{Q_A,Q_B} \big(\ket{\Phi^+} + \ket{\Psi^+}\big)
        + \sin\theta_\times\ket{1,1}_{R_B}\ket{V,V}_{Q_A,Q_B} \big(\ket{\Phi^+} - \ket{\Psi^+}\big)
    \bigg] \notag\\
\nonumber &\qquad
    + \frac{\sin\theta_\times}{2}\ket{1}_{R_E}
    \bigg[
        2\sin\theta_\times \ket{0,1}_{R_B}\ket{H,H}_{Q_A,Q_B}\ket{\Phi^+}
        \\
    &\qquad\qquad+ \cos\theta_\times\ket{1,1}_{R_B}\ket{H,V}_{Q_A,Q_B} \big(\ket{\Phi^+} + \ket{\Psi^+}\big)
        + \cos\theta_\times\ket{1,1}_{R_B}\ket{V,V}_{Q_A,Q_B} \big(\ket{\Phi^+} - \ket{\Psi^+}\big)
    \bigg]
\bigg) \\
\nonumber &+ \frac{1}{2}\ket{1,0}_{R_A}\bigg(\notag \frac{\sin\theta}{2}\ket{0}_{R_E}
    \bigg[
        2\sin\theta\, \ket{1,0}_{R_B}\ket{V,V}_{Q_A,Q_B}\ket{\Phi^+}
        \\
    &\qquad\qquad + \cos\theta\,\ket{0,0}_{R_B}\ket{V,H}_{Q_A,Q_B}(\ket{\Phi^+}+\ket{\Psi^+})
        + \cos\theta\,\ket{0,0}_{R_B}\ket{H,H}_{Q_A,Q_B}(\ket{\Phi^+}-\ket{\Psi^+})
    \bigg] \notag\\
\nonumber &\qquad
    + \frac{\cos\theta}{2}\ket{1}_{R_E}
    \bigg[
        2\cos\theta\, \ket{1,0}_{R_B}\ket{V,V}_{Q_A,Q_B}\ket{\Phi^+}
        \\
    &\qquad\qquad + \sin\theta\,\ket{0,0}_{R_B}\ket{V,H}_{Q_A,Q_B}(\ket{\Phi^+}+\ket{\Psi^+})
        + \sin\theta\,\ket{0,0}_{R_B}\ket{H,H}_{Q_A,Q_B}(\ket{\Phi^+}-\ket{\Psi^+})
    \bigg]
\bigg) \notag\\
\nonumber &+ \frac{1}{2}\ket{1,1}_{R_A}\bigg(\notag \frac{\sin\theta_\times}{2}\ket{0}_{R_E}
    \bigg[
        2\sin\theta_\times\, \ket{1,1}_{R_B}\ket{V,V}_{Q_A,Q_B}\ket{\Phi^+}
        \\
    &\qquad\qquad + \cos\theta_\times\ket{0,1}_{R_B}\ket{V,H}_{Q_A,Q_B}(\ket{\Phi^+}+\ket{\Psi^+})
        + \cos\theta_\times\ket{0,1}_{R_B}\ket{H,H}_{Q_A,Q_B}(\ket{\Phi^+}-\ket{\Psi^+})
    \bigg] \notag\\
\nonumber &\qquad
    + \frac{\cos\theta_\times}{2}\ket{1}_{R_E}
    \bigg[
        2\cos\theta_\times\, \ket{1,1}_{R_B}\ket{V,V}_{Q_A,Q_B}\ket{\Phi^+}
        \\
    &\qquad\qquad + \sin\theta_\times\ket{0,1}_{R_B}\ket{V,H}_{Q_A,Q_B}(\ket{\Phi^+}+\ket{\Psi^+})
        + \sin\theta_\times\ket{0,1}_{R_B}\ket{H,H}_{Q_A,Q_B}(\ket{\Phi^+}-\ket{\Psi^+})
    \bigg]
\bigg), \notag
\end{align}
}
following the conversion of the $Q_A, Q_B$ states into the $\{ \ket{H}\!, \ket{V}\}$ basis in the $\mu = 1$ case.

\section[\appendixname~\thesection]{Detection Probability}
\label{app:detProb}

In this Appendix, we derive the eavesdropper detection probability from the shared state obtained in Appendix~\ref{app:sharedstate}. We consider a fixed BB84 state specified by the bit value $b\in\{0,1\}$ and basis choice $\mu\in\{0,1\}$ prepared by \A.

For definiteness, consider first the state $\ket{H}$, corresponding to $(b,\mu)=(0,0)$. From Equation~\eqref{eq:app:sharedStateEve}, the component of the shared state corresponding to \E\ obtaining outcome $\ket{0}_{R_E}$ is
\begin{align}
    \ket{\Psi_{0_E}} = \frac{\cos\theta}{2} \Big[ & 2\cos\theta \ket{0,0}_{R_B} \ket{H,H} \ket{\Phi^+} \nonumber\\
    &+\sin\theta \ket{1,0}_{R_B} \ket{H,V} (\ket{\Phi^+}+\ket{\Psi^+}) \nonumber\\
    &+\sin\theta \ket{1,0}_{R_B} \ket{V,V} (\ket{\Phi^+}-\ket{\Psi^+}) \Big].
\end{align}
\E\ is detected whenever the control measurement projects onto the Bell state $\ket{\Psi^+}$, yielding the joint probability of detecting \E\ and obtaining \E's outcome $0$
\begin{equation}
    \mathbb{P}(\mathrm{detect},0_E|0,0) = \frac{1}{2}\cos^2\theta\,\sin^2\theta.
\end{equation}
The probability that \E\ obtains outcome $0$ is
\begin{equation}
    \mathbb{P}(0_E|0,0) = \cos^2\theta.
\end{equation}
Hence, the conditional detection probability is
\begin{equation}
    \mathbb{P}_{\mathrm{detect}|0,0,0} = \frac{ \mathbb{P}(\mathrm{detect},0_E|0,0) }{ \mathbb{P}(0_E|0,0) } = \frac{1}{2}\sin^2\theta.
\end{equation}
Repeating the same calculation for \E's orthogonal outcome gives
\begin{equation}
    \mathbb{P}_{\mathrm{detect}|1,0,0} = \frac{1}{2}\cos^2\theta.
\end{equation}

The same reasoning applies to any BB84 input state after replacing $\theta$ by the effective angle $\theta_{b,\mu}$ defined in Equation~\eqref{eq:thetaMu}. Thus,
\begin{align}
    \mathbb{P}_{\mathrm{detect}|0,b,\mu} &= \frac{1}{2}\sin^2\theta_{b,\mu}, \\ \mathbb{P}_{\mathrm{detect}|1,b,\mu} &= \frac{1}{2}\cos^2\theta_{b,\mu}.
\end{align}
The probabilities that \E\ obtains outcomes $0$ and $1$ are
\begin{equation}
    \mathbb{P}(0_E|b,\mu) = \cos^2\theta_{b,\mu}, \qquad \mathbb{P}(1_E|b,\mu) = \sin^2\theta_{b,\mu}.
\end{equation}
The total detection probability for the state specified by $(b,\mu)$ is therefore
\begin{align}
    p_{\mathrm{detect}|b,\mu} &= \mathbb{P}(0_E|b,\mu)\, \mathbb{P}_{\mathrm{detect}|0,b,\mu} + \mathbb{P}(1_E|b,\mu)\, \mathbb{P}_{\mathrm{detect}|1,b,\mu} \nonumber\\
    &= \cos^2\theta_{b,\mu}\sin^2\theta_{b,\mu}.
\end{align}
Using \eqref{eq:thetaMu}, the four BB84 preparations yield
\begin{align}
    p_{\mathrm{detect}|0,0} &= \cos^2\theta\sin^2\theta, \\
    p_{\mathrm{detect}|1,0} &= \cos^2\theta\sin^2\theta, \\
    p_{\mathrm{detect}|0,1} &= \cos^2\!\left(\theta-\frac{\pi}{4}\right) \sin^2\!\left(\theta-\frac{\pi}{4}\right), \\
    p_{\mathrm{detect}|1,1} &= \cos^2\!\left(\theta-\frac{\pi}{4}\right) \sin^2\!\left(\theta-\frac{\pi}{4}\right).
\end{align}
Using the trigonometric identity
\begin{equation}
    \cos^2\!\left(\theta-\frac{\pi}{4}\right) \sin^2\!\left(\theta-\frac{\pi}{4}\right) = \frac{1}{4} - \cos^2\theta\,\sin^2\theta,
\end{equation}
the last two expressions may also be written as
\begin{align}
    p_{\mathrm{detect}|0,1} &= \frac{1}{4} - \cos^2\theta\,\sin^2\theta, \\
    p_{\mathrm{detect}|1,1} &= \frac{1}{4} - \cos^2\theta\,\sin^2\theta.
\end{align}
Finally, averaging uniformly over the four BB84 states yields
\begin{align}
    p_{\mathrm{detect}} &= \frac{1}{4} \sum_{b,\mu \in \{0,1\}} p_{\mathrm{detect}|b,\mu} \nonumber\\
    &= \frac{1}{4} \left[ 2\cos^2\theta\sin^2\theta + 2\left( \frac{1}{4} - \cos^2\theta\sin^2\theta \right) \right] \nonumber\\
    &= \frac{1}{8},
\end{align}
which is independent of the polarizer angle $\theta$.

\section[\appendixname~\thesection]{Mutual Information}
\label{app:mutuinfo}

In this Appendix, we derive the probabilities entering the mutual-information calculation presented in Section~\ref{subsec:mutInfo}. These probabilities are obtained from the shared state derived in Appendix~\ref{app:sharedstate} after tracing over the polarization and path degrees of freedom, since the mutual information depends only on the classical registers associated with \A, \B, and \E.

Rather than working with the full expression of Equation~\eqref{eq:app:sharedStateEve}, it is convenient to rewrite the state in a compressed form. We absorb the joint polarization and path degrees of freedom into an orthonormal set of states $\{\ket{\ell_i}\}_{i=0}^{5}$, where the explicit form of the states is unimportant for the following derivation. Only their normalization and mutual orthogonality are required. The shared state can therefore be rewritten as
{
\begin{align}\label{eq:app:compressedState}
\ket{\Psi}_{\mathrm{shared}}^{(E)} &= \frac{1}{2}\ket{0,0}_{R_A}\bigg( \notag \cos\theta\ket{0}_{R_E}
    \bigg[
        \cos\theta \ket{0,0}_{R_B}\ket{0}_l
        + \frac{1}{\sqrt{2}}\sin\theta\,\ket{1,0}_{R_B}\ket{2}_l
        + \frac{1}{\sqrt{2}}\sin\theta\,\ket{1,0}_{R_B}\ket{4}_l
    \bigg] \notag\\
&\qquad +
    \sin\theta\ket{1}_{R_E}
    \bigg[
        \sin\theta \ket{0,0}_{R_B}\ket{0}_l
        + \frac{1}{\sqrt{2}}\cos\theta\,\ket{1,0}_{R_B}\ket{2}_l
        + \frac{1}{\sqrt{2}}\cos\theta\,\ket{1,0}_{R_B}\ket{4}_l
    \bigg]
\bigg) \notag\\
&+ \frac{1}{2}\ket{0,1}_{R_A}\bigg( \notag \cos\theta_\times\ket{0}_{R_E}
    \bigg[
        \cos\theta_\times \ket{0,1}_{R_B}\ket{0}_l
        + \frac{1}{\sqrt{2}}\sin\theta_\times\ket{1,1}_{R_B}\ket{2}_l
        + \frac{1}{\sqrt{2}}\sin\theta_\times\ket{1,1}_{R_B}\ket{4}_l
    \bigg] \notag\\
&\qquad +
    \sin\theta_\times\ket{1}_{R_E}
    \bigg[
        \sin\theta_\times \ket{0,1}_{R_B}\ket{0}_l
        + \frac{1}{\sqrt{2}}\cos\theta_\times\ket{1,1}_{R_B}\ket{2}_l
        + \frac{1}{\sqrt{2}}\cos\theta_\times\ket{1,1}_{R_B}\ket{4}_l
    \bigg]
\bigg) \\
&+ \frac{1}{2}\ket{1,0}_{R_A}\bigg(\notag \sin\theta\ket{0}_{R_E}
    \bigg[
        \sin\theta\, \ket{1,0}_{R_B}\ket{1}_l
        + \frac{1}{\sqrt{2}}\cos\theta\,\ket{0,0}_{R_B}\ket{3}_l
        + \frac{1}{\sqrt{2}}\cos\theta\,\ket{0,0}_{R_B}\ket{5}_l
    \bigg] \notag\\
&\qquad +
    \cos\theta\ket{1}_{R_E}
    \bigg[
        \cos\theta\, \ket{1,0}_{R_B}\ket{1}_l
        + \frac{1}{\sqrt{2}}\sin\theta\,\ket{0,0}_{R_B}\ket{3}_l
        + \frac{1}{\sqrt{2}}\sin\theta\,\ket{0,0}_{R_B}\ket{5}_l
    \bigg]
\bigg) \notag\\
&+ \frac{1}{2}\ket{1,1}_{R_A}\bigg(\notag \sin\theta_\times\ket{0}_{R_E}
    \bigg[
        \sin\theta_\times\, \ket{1,1}_{R_B}\ket{1}_l
        + \frac{1}{\sqrt{2}}\cos\theta_\times\ket{0,1}_{R_B}\ket{3}_l
        + \frac{1}{\sqrt{2}}\cos\theta_\times\ket{0,1}_{R_B}\ket{5}_l
    \bigg] \notag\\
&\qquad +
    \cos\theta_\times\ket{1}_{R_E}
    \bigg[
        \cos\theta_\times\, \ket{1,1}_{R_B}\ket{1}_l
        + \frac{1}{\sqrt{2}}\sin\theta_\times\ket{0,1}_{R_B}\ket{3}_l
        + \frac{1}{\sqrt{2}}\sin\theta_\times\ket{0,1}_{R_B}\ket{5}_l
    \bigg]
\bigg), \notag
\end{align}
}
where $\theta_\times = \theta - \tfrac{\pi}{4}$.

Since the states $\{\ket{\ell_i}\}$ are mutually orthonormal, tracing over the $\ell$ subsystem eliminates all coherence terms between different $\ket{\ell_i}$. The reduced density matrix of the classical registers is therefore diagonal, and the joint probabilities listed below are obtained directly from the squared amplitudes of the corresponding terms in Equation~\eqref{eq:app:compressedState}. We first compute the probabilities for a fixed choice of basis by \AB, corresponding to Figures~\ref{fig:mutual}a,b, before averaging over the four BB84 states to obtain the results shown in Figure~\ref{fig:mutual}c.

\subsection{Fixed-Basis Probabilities}

We first consider the case in which \AB\ both use the computational basis. Within this basis, the two possible key bits are assumed to be chosen uniformly. The marginal probabilities are therefore
\begin{equation}
    \mathbb{P}^{(Z)}(0_A) = \mathbb{P}^{(Z)}(1_A) = \mathbb{P}^{(Z)}(0_B) = \mathbb{P}^{(Z)}(1_B) = \mathbb{P}^{(Z)}(0_E) = \mathbb{P}^{(Z)}(1_E) = \frac{1}{2}.
\end{equation}

The joint probabilities for \AB\ are
\begin{align}
    \mathbb{P}^{(Z)}(0_A,0_B) &= \mathbb{P}^{(Z)}(1_A,1_B) = \frac{1}{2}(\cos^4\theta + \sin^4\theta), \\
    \mathbb{P}^{(Z)}(0_A,1_B) &= \mathbb{P}^{(Z)}(1_A,0_B) = \cos^2\theta\sin^2\theta.
\end{align}
Using
\begin{equation}\label{eq:app:fourthPow}
    \cos^4\theta + \sin^4\theta = 1 - 2\cos^2\theta\sin^2\theta,
\end{equation}
this becomes
\begin{align}
    \mathbb{P}^{(Z)}(0_A,0_B) &= \mathbb{P}^{(Z)}(1_A,1_B) = \frac{1}{2} - \cos^2\theta\sin^2\theta,\\
    \mathbb{P}^{(Z)}(0_A,1_B) &= \mathbb{P}^{(Z)}(1_A,0_B) = \cos^2\theta\sin^2\theta.
\end{align}
Similarly,
\begin{align}
    \mathbb{P}^{(Z)}(0_A,0_E) &= \mathbb{P}^{(Z)}(1_A,1_E) = \frac{1}{2}\cos^2\theta,\\
    \mathbb{P}^{(Z)}(0_A,1_E) &= \mathbb{P}^{(Z)}(1_A,0_E) = \frac{1}{2}\sin^2\theta,
\end{align}
and, by symmetry,
\begin{align}
    \mathbb{P}^{(Z)}(0_B,0_E) &= \mathbb{P}^{(Z)}(1_B,1_E) = \frac{1}{2}\cos^2\theta,\\
    \mathbb{P}^{(Z)}(0_B,1_E) &= \mathbb{P}^{(Z)}(1_B,0_E) = \frac{1}{2}\sin^2\theta.
\end{align}
The corresponding probabilities for the Hadamard basis are obtained by replacing $\theta \rightarrow \theta_\times$.

Figure~\ref{fig:p_all} shows the experimentally measured and theoretical probabilities entering the mutual-information calculation when \AB\ both choose the computational basis. These probabilities are used to evaluate the fixed-basis mutual information shown in Figure~\ref{fig:mutual}a.

\begin{figure}[t]
    \includegraphics[width=\linewidth]{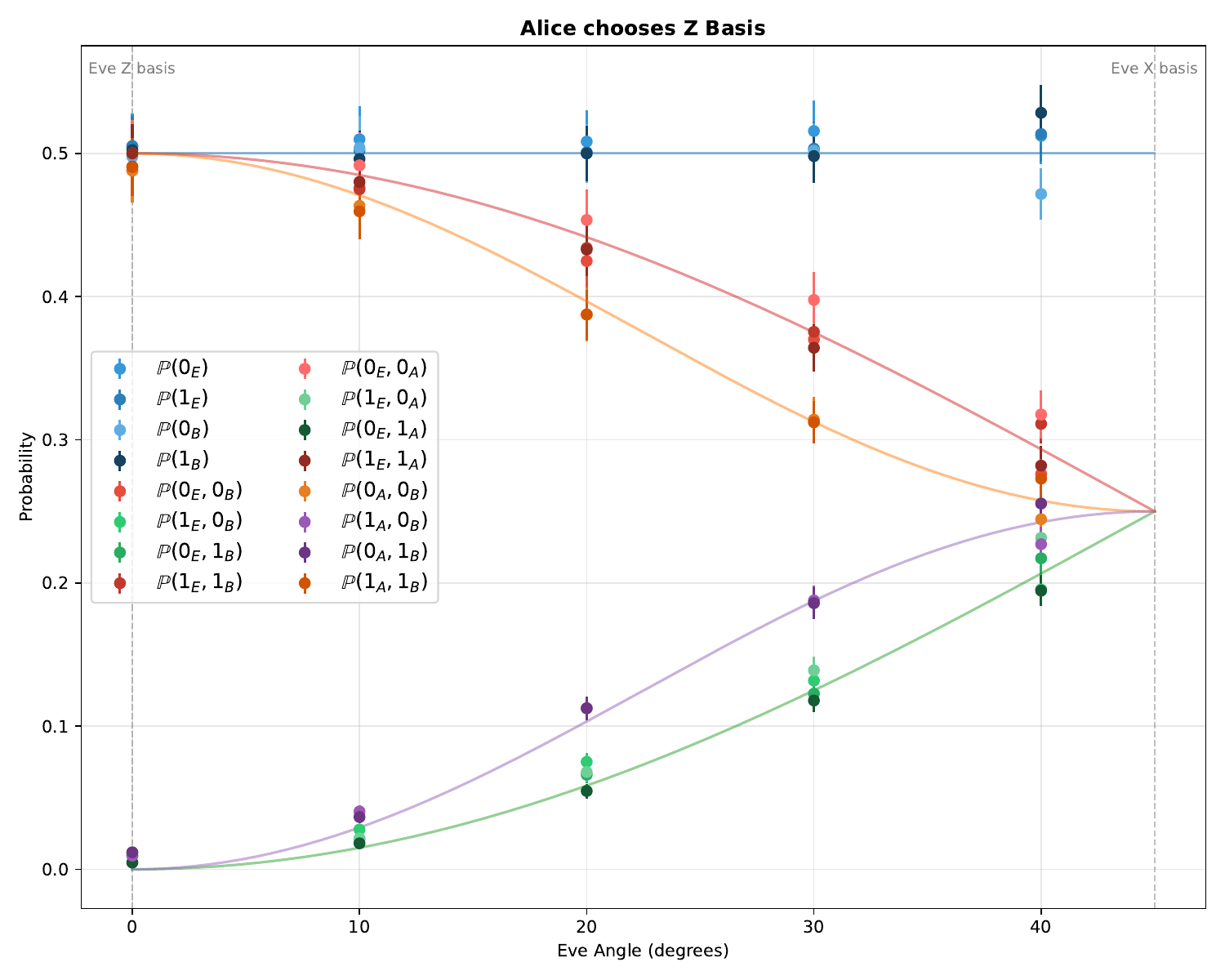}
    \caption{Measured and theoretical probabilities entering the mutual-information calculation when \AB\ both choose the computational basis. The solid curves are the theoretical predictions, while the markers are the experimental values. The plotted quantities include the marginal probabilities for \A, \B, and \E\ together with the relevant joint probabilities used to evaluate the mutual information.}
    \label{fig:p_all}
\end{figure}

\subsection{Average over the Four BB84 States}

We now consider the experimentally relevant situation in which \A\ chooses uniformly among the four BB84 states \eqref{eq:statesBB84}, and \E\ does not know this choice in advance. We average uniformly over the four BB84 preparations, i.e., over both basis choices and both bit values. Since \A\ chooses uniformly among the BB84 states, the single-register marginals remain uniform:
\begin{equation}\label{eq:uniformMarg}
    \mathbb{P}(0_A) = \mathbb{P}(1_A) = \mathbb{P}(0_B) = \mathbb{P}(1_B) = \mathbb{P}(0_E) = \mathbb{P}(1_E) = \frac{1}{2}.
\end{equation}
For \A\ and \E,
\begin{align}
    \mathbb{P}(0_A,0_E) &= \mathbb{P}(1_A,1_E) = \frac{1}{4}(\cos^2\theta + \cos^2\theta_\times),\\
    \mathbb{P}(0_A,1_E) &= \mathbb{P}(1_A,0_E) = \frac{1}{4}(\sin^2\theta + \sin^2\theta_\times).
\end{align}
By symmetry, the same expressions hold for \B\ and \E:
\begin{align}
    \mathbb{P}(0_B,0_E) &= \mathbb{P}(1_B,1_E) = \frac{1}{4}(\cos^2\theta + \cos^2\theta_\times),\\
    \mathbb{P}(0_B,1_E) &= \mathbb{P}(1_B,0_E) = \frac{1}{4}(\sin^2\theta + \sin^2\theta_\times).
\end{align}
For \AB,
\begin{align}
    \mathbb{P}(0_A,0_B) &= \frac{1}{4} \left( \cos^4\theta + \sin^4\theta + \cos^4\theta_\times + \sin^4\theta_\times \right),\\
    \mathbb{P}(0_A,1_B) &= \frac{1}{2} \left( \cos^2\theta \sin^2\theta + \cos^2\theta_\times \sin^2\theta_\times \right).
\end{align}
Using
\begin{equation}
    \cos^2\theta \sin^2\theta + \cos^2\theta_\times \sin^2\theta_\times = \frac{1}{4},
\end{equation}
together with \eqref{eq:app:fourthPow}, one finds
\begin{align}
    \mathbb{P}(0_A,0_B) &= \mathbb{P}(1_A,1_B) = \frac{3}{8},\\
    \mathbb{P}(0_A,1_B) &= \mathbb{P}(1_A,0_B) = \frac{1}{8}.
\end{align}

\subsection{Reduction of the Mutual Information}

The mutual information between two parties $X$ and $Y$ is defined in Equation~\eqref{eq:mutualInfo}. Both for the fixed-basis distributions and for the BB84-averaged distribution, the marginal probabilities are uniform and the joint probabilities satisfy
\begin{align}
    \mathbb{P}(0_X,0_Y) &= \mathbb{P}(1_X,1_Y), \\
    \mathbb{P}(0_X,1_Y) &= \mathbb{P}(1_X,0_Y),
\end{align}
together with normalization,
\begin{equation}
    \mathbb{P}(0_X,1_Y) = \frac{1}{2} - \mathbb{P}(0_X,0_Y).
\end{equation}
Substituting this into Equation~\eqref{eq:mutualInfo} gives
\begin{align}
    \mathcal{H}(X:Y) = 2 + 2\mathbb{P}(0_X,0_Y)\log_2\mathbb{P}(0_X,0_Y) + 2\mathbb{P}(0_X,1_Y)\log_2\mathbb{P}(0_X,1_Y),
\end{align}
for any $X,Y \in \{\A,\;\B,\; \E\}$. Introducing the bit agreement probability
\begin{equation}
    p_{X,Y} = \mathbb{P}(0_X,0_Y) + \mathbb{P}(1_X,1_Y) = 2\mathbb{P}(0_X,0_Y),
\end{equation}
one immediately obtains
\begin{equation}\label{eq:mutualCompactApp}
    \mathcal{H}(X:Y) = 1 + p_{X,Y}\log_2p_{X,Y} + (1-p_{X,Y})\log_2(1-p_{X,Y}).
\end{equation}

For the fixed computational basis ($\mu = 0$),
\begin{align}
    p_{A,B}^{(Z)} &= 1-2\cos^2\theta\sin^2\theta, \\
    p_{A,E}^{(Z)} = p_{B,E}^{(Z)} &= \cos^2\theta.
\end{align}
For the Hadamard basis ($\mu = 1$), the same expressions hold with $\theta$ replaced by $\theta_\times$. On the other hand, after averaging uniformly over the four BB84 states,
\begin{align}
    p_{A,B} &= 2\mathbb{P}(0_A,0_B) = \frac{3}{4}, \\
    p_{A,E} = p_{B,E} &= 2\mathbb{P}(0_A,0_E) = \frac{1}{2}(\cos^2\theta + \cos^2\theta_\times). \label{eq:app:agreeEve}
\end{align}
Therefore,
\begin{equation}
    \mathcal{H}(A:B) = 1 + \frac{3}{4}\log_2\frac{3}{4} + \frac{1}{4}\log_2\frac{1}{4} = -1 + \frac{3}{4}\log_2 3 \simeq 0.1887.
\end{equation}
Consequently, the averaged mutual information $\mathcal{H}(A:B)$ in Figure~\ref{fig:mutual}c is independent of $\theta$, even though the fixed-basis quantities in Figure~\ref{fig:mutual}a,b depend on $\theta$. The averaged agreement probability \eqref{eq:app:agreeEve} between \E\ and either legitimate party is maximal at $\theta = \frac{\pi}{8}$.

\subsection{Threshold Eavesdropping Strength}

When \E\ attacks only a fraction $t$ of the transmitted photons, we model the agreement probabilities as in Equation~\eqref{eq:strengthEve}. At the optimal angle $\theta = \pi/8$, the threshold at which \E\ obtains the same mutual information as \AB\ is determined by $\mathcal{H}(E:A/B) = \mathcal{H}(A:B)$. Since Equation~\eqref{eq:mutualCompactApp} is monotonic for agreement probabilities larger than $1/2$, this condition is equivalent to
\begin{equation}
    p_{E,A/B}(t) = p_{A,B}(t).
\end{equation}
At the optimal angle $\theta=\pi/8$, the relevant agreement probabilities are
\begin{align}
    p^{(\mathrm{no\ Eve})}_{E,A/B} &= \frac{1}{2}, \\
    p^{(\mathrm{with\ Eve})}_{E,A/B} &= \frac{1}{2}\left(1 + \frac{1}{\sqrt{2}}\right), \\
    p^{(\mathrm{no\ Eve})}_{A,B} &= 1, \\
    p^{(\mathrm{with\ Eve})}_{A,B} &= \frac{3}{4},
\end{align}
and we obtain
\begin{equation}
    (1-t)\frac{1}{2} + t\,\frac{1}{2} \left(1 + \frac{1}{\sqrt{2}}\right) = (1-t) + t\,\frac{3}{4}.
\end{equation}
Solving for $t$ gives
\begin{equation}
    t_0 = \frac{2}{1 + \sqrt{2}} \simeq 0.8284.
\end{equation}
Since the average detection probability is $p_{\mathrm{detect}} = t/8$, the threshold corresponds to
\begin{equation}
    p_{\mathrm{detect}} = \frac{1}{4(1 + \sqrt{2})} \simeq 0.1036.
\end{equation}

\end{widetext}

\end{document}